\pdfoutput=1
\documentclass[conference]{IEEEtran}
\IEEEoverridecommandlockouts
\usepackage{epsfig,endnotes}
\usepackage{color}
\usepackage{xspace}
\usepackage{url}
\usepackage{multirow}
\usepackage{makecell}
\usepackage{listings}
\usepackage{pifont}
\usepackage[]{algorithm2e}
\usepackage{comment}
\usepackage{hyperref}
\usepackage{tabularx}
\usepackage{flushend}
\let\labelindent\relax
\usepackage{enumitem}
\usepackage{subfig}
\usepackage{booktabs}
\usepackage{wrapfig}
\usepackage{color}
\usepackage{soul}
\usepackage{tikz} 
\usepackage{minibox}

\definecolor{lightgrey}{rgb}{0.925, 0.925, 0.925}
\sethlcolor{lightgrey}

\begin{document}

\date{}

\newcommand{\bheading}[1]{{\vspace{4pt}\noindent{\textbf{#1}}}}
\newcommand{\iheading}[1]{{\vspace{4pt}\noindent{\textit{#1}}}} 

\newcommand{\draft}[1]{\textcolor{blue}{#1}}

\newcommand{\etal}{\emph{et al.}\xspace}
\newcommand{\etc}{\emph{etc}\xspace}
\newcommand{\ie}{\emph{i.e.}\xspace}
\newcommand{\eg}{\emph{e.g.}\xspace}
\newcommand{\aka}{\emph{a.k.a.}\xspace}

\newcommand{\figurewidth}{\columnwidth}
\newcommand{\secref}[1]{\mbox{Sec.~\ref{#1}}\xspace}
\newcommand{\secrefs}[2]{\mbox{Sec.~\ref{#1}--\ref{#2}}\xspace}
\newcommand{\figref}[1]{\mbox{Fig.~\ref{#1}}}
\newcommand{\tabref}[1]{\mbox{Table~\ref{#1}}}
\newcommand{\lstref}[1]{\mbox{Listing~\ref{#1}}}
\newcommand{\appref}[1]{\mbox{Appendix~\ref{#1}}}
\newcommand{\ignore}[1]{}

\newcommand{\sysname}{\textsc{SpeechMiner}\xspace}

\newcommand{\pluginonename}{{instruction sequence}\xspace}
\newcommand{\plugintwoname}{{Analysis Module Plugin}\xspace}

\newcommand{\uop}{\ensuremath{\mu}op\xspace}
\newcommand{\uops}{\ensuremath{\mu}ops\xspace}
\newcommand{\vulname}{\textsc{Speech}\xspace}
\newcommand{\phaseone}{\textsc{P1}\xspace}
\newcommand{\phasetwo}{\textsc{P2}\xspace}
\newcommand{\atkname}{{speculative execution attack}\xspace}
\newcommand{\atknames}{{speculative execution attacks}\xspace}

\newcommand{\SP}{Speculation Primitive}

\newcommand{\flushreload}{\textsc{Flush-Reload}\xspace}
\newcommand{\Flush}{\textsc{Flush}\xspace}
\newcommand{\Reload}{\textsc{Reload}\xspace}
\newcommand{\primeprobe}{\textsc{Prime-Probe}\xspace}
\newcommand{\Prime}{\textsc{Prime}\xspace}
\newcommand{\Probe}{\textsc{Probe}\xspace}
\newcommand{\evictreload}{\textsc{Evict-Reload}\xspace}
\newcommand{\flushflush}{\textsc{Flush-Flush}\xspace}
\newcommand{\flushfunc}{\texttt{clearcache}\xspace}

\newcommand{\gbytes}{\ensuremath{\mathrm{GB}}\xspace}
\newcommand{\mbytes}{\ensuremath{\mathrm{MB}}\xspace}
\newcommand{\kbytes}{\ensuremath{\mathrm{KB}}\xspace}
\newcommand{\bytes}{\ensuremath{\mathrm{B}}\xspace}
\newcommand{\hertz}{\ensuremath{\mathrm{Hz}}\xspace}
\newcommand{\ghertz}{\ensuremath{\mathrm{GHz}}\xspace}
\newcommand{\msecs}{\ensuremath{\mathrm{ms}}\xspace}
\newcommand{\usecs}{\ensuremath{\mathrm{\mu{}s}}\xspace}
\newcommand{\nsecs}{\ensuremath{\mathrm{ns}}\xspace}
\newcommand{\secs}{\ensuremath{\mathrm{s}}\xspace}
\newcommand{\gbits}{\ensuremath{\mathrm{Gb}}\xspace}

\newcommand\yz[1]{\textcolor{red}{\{\textbf{yinqian:} {\em#1}\}}}

\newcommand\yx[1]{\textcolor{blue}{\{\textbf{yuan:} {\em#1}\}}}

\newcounter{packednmbr}
\newenvironment{packedenumerate}{
\begin{list}{\thepackednmbr.}{\usecounter{packednmbr}
\setlength{\itemsep}{0pt}
\addtolength{\labelwidth}{4pt}
\setlength{\leftmargin}{12pt}
\setlength{\listparindent}{\parindent}
\setlength{\parsep}{3pt}
\setlength{\topsep}{3pt}}}{\end{list}}

\newenvironment{packeditemize}{
\begin{list}{$\bullet$}{
\setlength{\labelwidth}{8pt}
\setlength{\itemsep}{0pt}
\setlength{\leftmargin}{\labelwidth}
\addtolength{\leftmargin}{\labelsep}
\setlength{\parindent}{0pt}
\setlength{\listparindent}{\parindent}
\setlength{\parsep}{2pt}
\setlength{\topsep}{1pt}}}{\end{list}}

\newcommand{\cmark}{\ding{51}}%
\newcommand{\xmark}{\ding{55}}%

\lstdefinelanguage
   [x64]{Assembler}     
   [x86masm]{Assembler} 
   {morekeywords={CDQE,CQO,CMPSQ,CMPXCHG16B,JRCXZ,LODSQ,MOVSXD, %
                  POPFQ,PUSHFQ,SCASQ,STOSQ,IRETQ,RDTSCP,SWAPGS, %
                  rax,rdx,rcx,rbx,rsi,rdi,rsp,rbp, %
                  r8,r8d,r8w,r8b,r9,r9d,r9w,r9b,r10,r11,r12,enclu,mfence,lfence,cmova,cmovg,cmovl}}
                  
\definecolor{auburn}{rgb}{0.43, 0.21, 0.1}
\definecolor{codegreen}{rgb}{0,0.6,0}
\definecolor{codegray}{rgb}{0.5,0.5,0.5}
\definecolor{codepurple}{rgb}{0.58,0,0.82}
 
\lstdefinestyle{mystyle}{
	language=[x64]Assembler,
    backgroundcolor=\color{white},   
    commentstyle=\color{blue},
    keywordstyle=\color{black},
    numberstyle=\tiny\color{black},
    stringstyle=\color{black},
    basicstyle=\scriptsize,
    breakatwhitespace=false,         
    breaklines=true,                 
    captionpos=b,                    
    keepspaces=true,                 
    numbers=left,                    
    numbersep=5pt,                  
    showspaces=false,
    xleftmargin=0.25cm,
    xrightmargin=0.25cm,
    showstringspaces=false,
    showtabs=false,                  
    tabsize=2
}
\lstset{
  style=mystyle
}



\title{\sysname: A Framework for Investigating and Measuring Speculative Execution Vulnerabilities}


\author{Yuan Xiao, Yinqian Zhang, Radu Teodorescu\\
Department of Computer Science and Engineering\\
The Ohio State University\\
\{xiao.465\}@osu.edu \hfill \{yinqian, teodores\}@cse.ohio-state.edu}

\maketitle
\thispagestyle{plain}
\pagestyle{plain}


\setlength{\intextsep}{2pt} 
\setlength{\textfloatsep}{3pt} 
\setlength{\floatsep}{3pt}

\captionsetup{font=small}

\lstset{
commentstyle=\color{green},
numbers=left,
numberstyle=\tiny,
numbersep=5pt,
language={[x86masm]Assembler},
frame = single}

\begin{abstract}

SPEculative Execution side Channel Hardware (\vulname) Vulnerabilities have enabled the notorious Meltdown, Spectre, and L1 terminal fault (L1TF) attacks. While a number of studies have reported different variants of \vulname vulnerabilities, they are still not well understood. This is primarily due to the lack of information about microprocessor implementation details that impact the timing and order of various micro-architectural events. Moreover, to date, there is no systematic approach to quantitatively measure \vulname vulnerabilities on commodity processors. 

This paper introduces \sysname, a software framework for exploring and measuring \vulname vulnerabilities in an automated manner. \sysname empirically establishes the link between a novel two-phase fault handling model and the exploitability and speculation windows of \vulname vulnerabilities. It enables testing of a comprehensive list of exception-triggering instructions under the same software framework, which leverages covert-channel techniques and differential tests to gain visibility into the micro-architectural state changes. We evaluated \sysname on 9 different processor types, examined 21 potential vulnerability variants, confirmed various known attacks, and identified several new variants.

\end{abstract}

\section{Introduction}

Speculative  Execution  Side  Channel  Hardware  Vulnerabilities~\cite{msftblogspeculative} are computer micro-architectural vulnerabilities in modern pipelined processors that, due to speculative and out-of-order execution, may execute instruction sequences that should not be executed if instructions are strictly executed in program order. Speculatively executed instructions may lead to information leakage as they lead to state changes in cache in the same way as retired instructions. Such vulnerabilities are the root causes of the well-known Meltdown~\cite{Lipp2018meltdown}, Spectre~\cite{Kocher2018spectre}, Foreshadow~\cite{vanbulck2018foreshadow},  and RIDL~\cite{van2019ridl}. 


Although these security attacks are high-profile due to their severe consequences, they are unlikely to be completely eliminated in modern high-performance processors, because transient execution (including speculative and out-of-order execution), implicit caching, and aggressive prefetching offer significant performance gains. While some of these vulnerabilities can be mitigated by microcode patches or hardware fixes~\cite{intelvariant3a,amdvariant3a,armvariant3a}, others have to be temporarily mitigated by software~\cite{msftblogspeculative,linuxpatch,armpatch,llvmpatch}. Moreover, new variants of these vulnerabilities are constantly being discovered by hackers and security researchers. Prominent examples include LazyFP~\cite{stecklina2018lazyfp}, Meltdown-RW~\cite{kiriansky2018speculative}, Fallout~\cite{fallout}, ZombieLoad~\cite{schwarz2019zombieload}, \etc.

A major challenge faced by researchers, software developers and hardware designers is the ignorance about the fundamental question of what determines the success or failure of an attack. 
Without a concrete general conclusion over the nature of these attacks, great efforts are put into figuring out unique mitigation for each newly-emerging variant. Evaluating variants is also difficult with only random attempts of seemingly relative implementation tricks, hoping for a successful exploitation.
Three aspects of complexity lead to the difficulty for a general conclusion to be made. 
First, the attacks vary greatly from each other. They have different threat models and exploit different instructions.
And not enough details are provided about their implementation.
Second, the micro-architectural states during execution is unobservable and unpredictable. The aggressive speculative and out-of-order processor workflow leads to great complication for the execution of even one instruction.
Third, the design and implementation of computer micro-architectures are highly variable depending on processor generation and manufacturer. The same variant of a vulnerability may manifest on one processor family but not others. Therefore, given a commodity processor, there is no method that could affirmatively assert that a specific processor is free of all known vulnerabilities: The only result that can be demonstrated by security researchers is a successful attack on a particular CPU under a certain condition, but unsuccessful attempts do not offer a sense of security.
The goal of the paper is to (1) comprehensively understand the SPEculative Execution side Channel Hardware (\vulname)\footnote{Speculative Execution Side Channel Hardware Vulnerability is the term preferred by Intel and Microsoft. We omit ``S'' in the acronym to reflect the debate of whether a side channel or a covert channel is used in such attacks.} Vulnerabilities~\cite{msftblogspeculative} in modern computer micro-architectures and (2) systematically and quantitatively evaluate \vulname vulnerabilities on commodity processors, including providing deterministic evidence for inexploitable variants. 
As the Spectre-type vulnerabilities~\cite{canella2018systematic} are due to illegal poisoning of branch prediction rather than hardware implementation flaws in the prediction units themselves, we put emphasis on the more complex Meltdown-type \vulname vulnerabilities caused by fault handling. All Meltdown-type vulnerabilities~\cite{canella2018systematic} require such faults that may or may not trigger an explicit exception. In this paper, we intend to get a comprehensive understanding of the general Meltdown-type vulnerabilities. Thus, we focus on the core x86 ISA but leave hardware extensions such as SGX and VMX to future work. 

However, it is very challenging to precisely determine the internal micro-architectural implementation of the processors, which is not made public in sufficient detail by the processor vendors. Moreover, while some of the micro-architectural design choices may be available, implementation details such as timing and order of events are not documented. To achieve the first research goal, we propose a novel two-phase model to describe the execution of x86 instructions with respect to the handling of faults. The model abstracts away complex implementation details and focuses on the software-observable and measurable events that are relevant to \vulname vulnerabilities. More specifically, the two-phase model describes the exploitation of a \vulname vulnerability as the outcome of two race conditions: A race condition between data fetching and processor fault handling and a race condition between covert channel transmission and speculative instruction squashing. The \textit{exploitability} and \textit{speculation window} can be determined by these two race conditions.



To achieve the second research goal, we designed a software framework, dubbed \sysname, to systematically and quantitatively measure the exploitability and the speculation windows of a variety of \vulname vulnerabilities. However, building such an analytical framework is technically challenging. \sysname brings forward the following solutions to this non-trivial task:

To gain visibility into the micro-architecture, \sysname employs covert-channel techniques to indirectly infer micro-architectural state changes. To establish the link between the two-phase model and the \vulname vulnerabilities, \sysname incorporated several carefully designed experiments to infer the internal implementations of the tested computer micro-architecture. To enable quantitative analysis, \sysname dynamically adjusts the tested instruction sequences and utilizes differential tests to quantify the exploitability and speculation windows.  Finally, to enable systematic analysis, instead of exhausting all micro-architectural uses of speculative execution, we enumerate all architecturally-observable exceptions by referencing the software development manuals from the vendors (though in a manual way) to instantiate the test cases of \sysname.

We run \sysname on 9 different types of processors to examine 21 different variants of \vulname vulnerabilities. our experiments not only confirmed previously demonstrated attacks, but also identified a few new exploitable variants on the tested Intel and AMD machines, which have been reported to the vendors. Moreover, \sysname enabled us to perform quantitative measurements of the exploitability and speculation windows of these vulnerabilities. The significance of the quantitative analysis is that it provides security assurance for the negative results---a processor not vulnerable in one of our exploitability tests is assured to be immune from the corresponding attack.

Moreover, our study 
yields some very interesting discoveries. For instance, it explains why zero values are sometimes returned by the Meltdown-US attacks; it suggests that the speculation window of any faulting instructions can be controlled and tuned by the attacker; it rules out the possibility of advanced attacks by nesting multiple speculative instructions; it explains why Meltdown-US attacks can leak data not present in the L1 cache, but L1TF attacks cannot.


In summary, this work makes the following contributions:

\begin{packeditemize}
\item It proposes a novel two-phase model to describe x86 fault handling and its relationship to the exploitability and speculation windows of \vulname vulnerabilities.

\item It designs and implements \sysname framework to automatically and systematically explore and measure \vulname vulnerabilities. 

\item It enables quantitative measurements of the exploitability of \vulname vulnerabilities, providing security assurance to negative test results.

\item It explains the root causes of some observations made by prior studies and clarifies common misunderstandings.

\item It performs automated tests of 21 vulnerability variants on 9 processor types; it confirms existing vulnerabilities and uncovers a few new variants of \vulname vulnerabilities. 
\end{packeditemize}

\section{Modeling \vulname Vulnerabilities}
\label{sec:model}



\subsection{Documented Instruction Execution Model}
\label{sec:method:model}


\begin{figure*}[t]
\centering
\includegraphics[width=2\columnwidth]{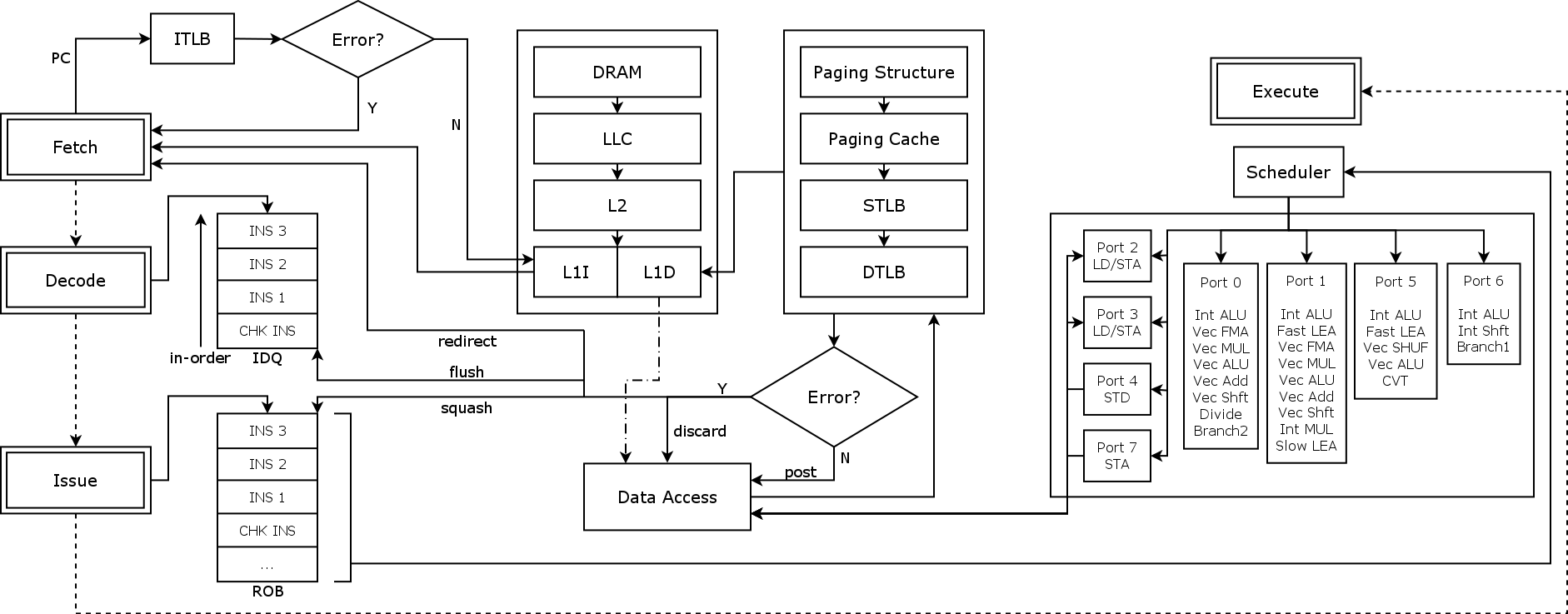}
\caption{Instruction execution model of x86 (illustrated using Skylake processors).}
\label{fig:model}
\end{figure*}

Although the exact internal implementation of a commodity processor is proprietary to each processor vendor, some design details are made available in Intel and AMD software developer manuals, white papers, patent applications, as well as technical blogs written by computer architects and hardware engineers. The execution model we build in this paper abstracts away aspects that are not relevant to \vulname vulnerabilities. We show an overview of typical out-of-order execution engine in \figref{fig:model}. As the figure shows, instruction execution follows five main stages: instruction fetch, decode, issue, execute (including memory access) and retire. 




\subsubsection{Fetching, Decoding, Execution and Retirement}

We first model the five stages for an instruction in the execution engine.

\bheading{Instruction fetching.}
The front end of the processor fetches instructions from the L1 instruction cache. Since instruction addresses are virtual, they  must first be translated into physical addresses before a fetch request can be sent to memory. Virtual to physical address translations are cached in the instruction TLB (ITLB). If a translation is not present in the ITLB, a look-up request will be sent to the second-level TLB (STLB), paging structure caches, or the page tables in the memory. 

An access permission check is performed simultaneously with the address translation. If the check fails (\eg due to an illegal address), the front end will immediately raise an exception and start fetching instructions from the exception handler. The TLB entry may also be marked as invalid. The instruction at the illegal address will not be decoded or issued for execution~\cite[Chapter 2.3.2]{inteloptimization}. These instructions therefore cannot serve as the basis for a \atkname. If the permission check succeeds, instructions are fetched from the L1 Instruction cache, lower-level caches, or the memory. They are then passed on to the decode stage. 




\bheading{Instruction decoding.} The instruction decoder is responsible for interpreting the instruction, identifying operands and, in the case of x86, translating the complex (CISC) instructions into a simpler internal representation called micro-operations (\aka, \uops). \uops are not visible to the programmer and follow the reduced instruction set (RISC) design. This means they use two input operands and one output, all arithmetic and logic operations are performed on register operands and the only instructions that access memory are Loads and Stores. Decoded instructions are added to the Instruction Decode Queue (IDQ) in program order. This marks the end of the front-end of the processor and the last step in which instructions are processed in program order. 


\bheading{Instruction issuing.} From the IDQ, \uops are issued in FIFO order to the back end of the pipeline. Once issued, they are no longer constrained by program order and can execute out-of-order, as soon as their operands are available. While \uops can execute out-of-order, they are required to commit their results to the visible processor state (architectural state) in program order, to preserve correctness. A hardware structure called a reorder buffer (ROB) is used to keep track of the \uops program order, while they are in the back-end of the pipeline. The ROB is a table that records all \uops in execution and their associated status (\eg operands pending, in-execution, completed, etc.) When issued, \uops are added to the ROB, in FIFO order, as long as there are available slots. 


\bheading{Instruction execution and retirement.} All \uops operands are renamed and their dependencies tracked with the help of hardware structures called reservation stations. Once issued, \uops are eligible for execution provided that their operands are ready and execution resources are available. \uops are executed out-of-order and in parallel. However, if a \uop has data dependency on its preceding \uops, it has to wait until the dependency is resolved before being scheduled for execution. When all conditions are met \uops are dispatched to the appropriate execution units through hardware structures called ports. Multi-cycle operations can occupy execution units, possibly stalling other \uops demanding the same resources. 

When \uops finish execution they write back their results to so-called physical registers that are not part of the architectural state and are not visible to the program. Results are also forwarded to dependent \uops through dedicated bypass data paths allowing dependent \uops to be scheduled for execution in the same cycle. 

While \uops can execute out-of-order, they are required to commit their results to architectural state visible to the program (including architectural registers and memory) in program order. The ROB is used to enforce this requirement by committing and retiring \uops in FIFO order. As \uops of an instruction reach the head of the ROB, if they have all finished execution, they can commit their results and retire, at which point they are removed from the ROB. 

Transient execution relies on hardware to prevent transient instructions from changing the architectural state visible to the program, until instructions are determined to be correct. When mis-speculation is detected (\eg a branch is mis-predicted or an exception is triggered), all mis-speculated instructions have to be squashed. Precise handling of transient state requires that all instructions that precede the first mis-speculated instruction must commit, and all other mis-speculated instructions must be squashed. 



\subsubsection{Memory Accesses and Address Translation}

The \uops that perform memory accesses are executed in specific execution units. In 32-bit mode, the logical address is first translated into linear address by referencing the segment descriptor; in 64-bit mode, the logical address is the same as the linear address. Then given the linear address of the data in memory, TLB is first consulted to look for the physical address. If the corresponding entry is not available in the data TLB (DTLB), the secondary TLB (STLB) is searched. Similarly, the paging structure cache and page tables in memory are looked up if an STLB miss is encountered. When walking the page tables, the corresponding page directory entries are loaded into the paging structure cache. 
When the page table entry (PTE) for the 4KB page is eventually located, it will be inserted into STLB and DTLB. 
Given a physical address from the DTLB, the data fetching starts from the L1 cache which is the fastest in the memory subsystem. Should there be a L1 miss, the L2 cache, the last-level cache (LLC), and the DRAM are checked one by one until there is a hit. Upon a hit, the data will be pushed to all levels of cache and L1 will pass it to the execution units. 
Processor internal buffers such as Data Cache Unit (DCU), Line Fill Buffer (LFB), Load Buffer (LB) and Store Buffer (SB), as components of L1 cache, are not paid enough attention to by the security community until recent disclosure of transient execution attacks leveraging them \cite{van2019ridl,fallout,schwarz2019zombieload}. 

\subsection{Detecting and Handling Mis-Speculation}
\label{subsec:handling}

Although high-level information on instruction execution is documented in the manuals of vendors, the internal fault handling during speculative execution (including out-of-order execution) remains unclear and is complex due to aggressive out-of-order implementation. We propose a two-phase model to understand the internal micro-architectural implementation of fault handling. Our model abstracts the unnecessary, maybe unclear, hardware implementation details into logical events that are directly related to \vulname. The two phases to be explained not only clarify the exact fault handling scheme but also correspond to the two race conditions faced by an attack. This offers us an opportunity to study them separately and understand them systematically. The model will be later validated through experiments described in \secref{sec:understanding}. 


\begin{figure}[htp]
\centering
\subfloat[Two phases of fault handling.]{
    \includegraphics[clip,width=0.9\columnwidth]{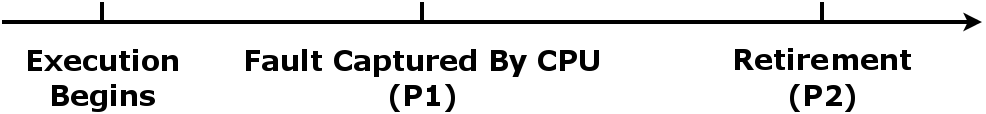}
    \label{subfig:two-phase}
}

\subfloat[Exploitable fault handling.]{
    \includegraphics[clip,width=0.9\columnwidth]{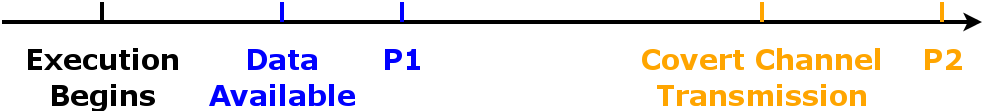}
    \label{subfig:exploitable}
}

\subfloat[Non-exploitable fault handling.]{
    \includegraphics[clip,width=0.9\columnwidth]{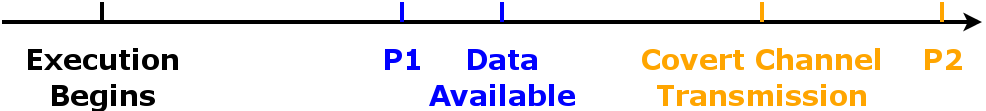}
    \label{subfig:inexploitable}
}

\caption{The two-phase model for fault handling.}
\label{fig:two-phase}
\end{figure}

The security check starts at the same time as data fetching, but proceeds asynchronously. If the check passes, the fetched data can be finally committed. Otherwise, the processor will handle the fault (\eg, by raising an exception) and clean up the pipeline by discarding the execution results and squashing \uops not supposed to execute.
As shown in \figref{subfig:two-phase}, the exceptions are handled in two phases: In the first phase (dubbed \phaseone), defined as when the processor detects an error in the \uop, the exception is passed to the corresponding execution unit immediately, which reacts to it by stopping the execution of the \uop. If the \uop performs a data fetching, \textit{two} cases may happen: If the data is not yet retrieved, the fetching will be suspended and a dummy value (\eg, zero) is returned as the data, as in \figref{subfig:inexploitable}. If the data loading has already finished at that time, it will not be affected. The data fetching is immediately forwarded to \uops in the ROB that are waiting for it. This is demonstrated in \figref{subfig:exploitable}. It in fact describes one of the two race conditions for a \vulname attack to succeed, between speculative data fetching and processor fault handling. 


The second phase of exception handling (dubbed \phasetwo) happens when the faulting \uop reaches the head of ROB and is ready to retire. The processor checks any pending exceptions with the \uops of the retiring instruction and, if detected, the entire execution engine is cleansed in the following way. First, all following \uops in the ROB are squashed: already executed \uops will never retire and their execution results are discarded; \uops that are not yet executed will not be executed any more. Second, the IDQ will stop issuing more decoded \uops to ROB. IDQ will be flushed for optimal future performance. 
Third, information about the exception is saved in relevant registers. Lastly, the front end will be redirected to exception handler. After all preparation is done by the processor, the exception handler of the OS will take over the control of the CPU. This leads to the other race condition for the attack. The covert channel transmission instructions executed speculatively should conclude before \phasetwo. Otherwise, they are going to be squashed and never get a chance to transfer the secret to the attacker.

\section{\sysname Framework}
\label{sec:framework}



To explore \vulname vulnerabilities of a processor in an automated (or semi-automated) manner, we designed and implemented a software framework, dubbed \sysname, in which a sequence of x86 instructions is constructed using templates and executed in a controlled environment. The \sysname framework provides the users with interfaces to select the instruction sequences for testing and to analyze the results of the tests. The user can use scripting languages to automate large-scale tests using these interfaces.

As the same instruction sequence may exhibit non-deterministic behaviors at the micro-architectural level when executed with different micro-architectural conditions (\eg, cache and TLB conditions, memory bus status, \etc.), the \sysname framework is designed to provide control and abstraction of these external conditions. Moreover, the framework is designed to work in both kernel mode and user mode, allowing tests of instruction sequences that operate in both modes; it also supports both 32-bit and 64-bit architectures.


\begin{figure}[t]
\centering
\includegraphics[width=\columnwidth]{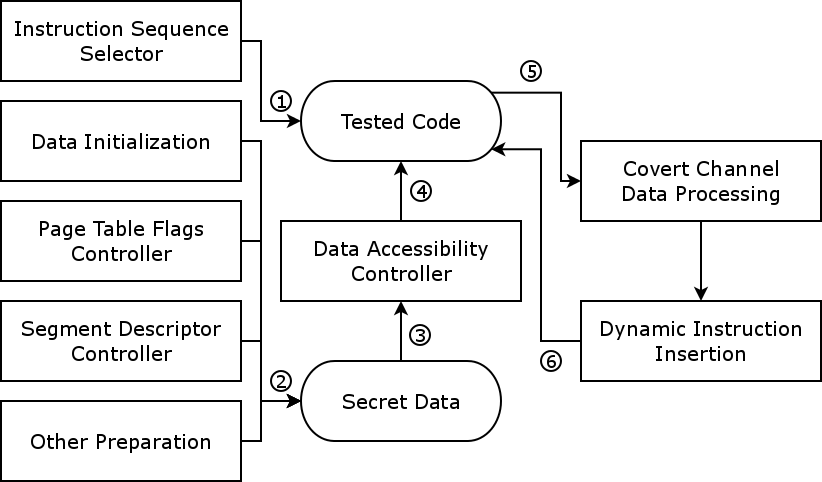}
\caption{Architecture and workflow of \sysname.}
\label{fig:framework}
\end{figure}


\subsection{Architecture of \sysname}
\label{subsec:architecture}

\figref{fig:framework} illustrates the architecture and workflows of the \sysname framework. 
Rectangles represent software components and ellipses represent code/data. \sysname consists of the following components. The \textit{Instruction Sequence Selector} of \sysname selects the instruction sequence for testing a certain variant of \vulname vulnerability. Then a dummy secret data is initialized in the memory and the required page table flags, segment descriptor or other settings are configured. Next, \sysname starts to run experiments. In each round of the test, the \textit{Data Accessibility Controller} module first sets the desired execution environments such as the status of caches and TLBs. Then the experiment is conducted and raw output data is collected via covert channels. The \textit{Covert Channel Data Processing} module analyzes the raw data, generating either the final analysis results or instructing the \textit{Dynamic Instruction Insertion module} to modify the tested instruction sequence for the next round of experiments.


\bheading{Instruction Sequence Selector.}
In each test, one instruction sequence is selected by the Instruction Sequence Selector. 


\bheading{Secret Data Initialization.}
In many of the tests, the secret value to be extracted is stored in the memory. The \sysname framework simulates the targeted secret using a 64-bit integer variable (in 64-bit mode) and initializes it to be a specified value (\eg, 0x42000 as used in following code examples). A single MOVQ instruction could load this secret value from the memory into a register. The size of the secret is reduced to 32 bits in the 32-bit mode tests. 

\bheading{Page Table Flag Controller.}
For cases that require modification of page table entries (PTE), we implemented a kernel module that allows setting or clearing specific PTE flags.

\bheading{Segment Descriptor Controller.}
Segmentation is still implemented in all modern x86 processors. When running the tool in the 32-bit mode, segmentation is enabled. Segment Descriptor Controller is provided to generate required segment descriptors in the Local Descriptor Table to trigger exceptions by violating segmentation-related rules.

\bheading{Other Preparation.}
Besides these common preparation components, in some cases, \sysname also needs to take care of some special needs such as configuring memory protection keys~\cite{intelmpk}, enabling SMAP~\cite{smap}, \etc.

\bheading{Data Accessibility Controller.}
To ensure a deterministic execution environment, the \sysname framework needs to control the status of the cache copies of specific memory blocks and the TLB entries of a specific memory page. The Data Accessibility Controller utilizes preloading and flushing techniques to control TLB and cache entries. Some technical challenges may arise, however.

The preloading of TLB entries may trigger exceptions (which is expected in our design). To preload the TLB entry of a kernel page, the Data Accessibility Controller preloads its TLB entry in a kernel module. To preload TLB entries with \texttt{Reserved} flag set or \texttt{Present} flag cleared, directly loading the corresponding page can preload entries in the paging structure caches~\cite[Chapter~4.10]{IntelDevelopmentManual}. While valid TLB entries may not be created, invalid entries may be created.  
Flushing TLB entries are performed in the supervisor model via a kernel module. One method is to leverage the INVLPG instruction which flushes one single TLB entry and the related paging structure cache entries. The other is to reload the CR3 register which will flush all TLB entries and the whole paging structure cache.

Preloading and flushing the cache entries of a data block are performed on its shadow virtual addresses. For each data block that needs fine-grained control of its cache status (\ie, on which cache level a copy is presented), two different virtual address mappings are provided for the same memory page that stores the data block: One mapping is used by the test that triggers exceptions, while the other is used as the shadow virtual address that does not block accesses. To force data in L1, it is directly preloaded via the shadow address. To make the data block in L2 (but not in L1), it needs to be preloaded first and then evicted from L1 using an eviction set~\cite{liu2015last}. As the LLC cache is shared among all physical cores, after flushing the data to memory (using the CLFLUSH instruction), preloading it from another physical core ensures that the data resides in LLC but not in the L1 or L2 caches (of the tested core). Moreover, as preloading or flushing cache entries also preloads the TLB entry of the page, additional procedures must be taken if this side effect is undesired.

\bheading{Covert Channel Data Processing.} The covert channel signals collected during the test are processed and analyzed. If needed, it provides feedback to the Dynamic Instruction Insertion module to repeat the test with adjusted the instruction sequences to be tested. 


\bheading{Dynamic Instruction Insertion.}
The module is implemented by altering the code at runtime. It dynamically adjusts a given code sequence according to the need of the experiments (\eg by inserting a certain number of ADD/SUB instructions).


\bheading{Handling or Suppressing Exceptions.} As the tested instruction sequences may trigger exceptions, the framework must handle or suppress exceptions properly. When executed in user mode, exceptions are dealt within signal handlers to ensure compatibility on all hardware; when executed in supervisor mode, the exceptions are suppressed using Retpoline \cite{retpoline}, as is done by Stecklina \etal~\cite{stecklina2018lazyfp}.


\subsection{Instruction Sequences}
\label{subsec:exception_plugin}



To trigger different types of faults, the tested instruction sequences may have distinct structures. Nevertheless, we managed to build a uniform modular template for all tested instruction sequences. Specifically, a template consists of three components: a Windowing Gadget, a Speculation Primitive, and a Disclosure Gadget\footnote{The terminologies follow the suggestions from Intel and Microsoft~\cite{msftblogspeculative}.}. 


\bheading{Speculation Primitives.} A Speculation Primitive consists of one or two instructions that will trigger a fault when executed. 


\bheading{Windowing Gadgets.} A Windowing Gadget consists of a sequence of instructions that precedes the Speculation Primitive. It serves two purposes: to enlarge the speculation window and to eliminate side-effects of instruction issuing. These two purposes can be satisfied by delaying the retirement of the Speculation Primitive, which can be achieved by three means: (1) Delaying the retirement of the instructions of the Windowing Gadget. This is because an instruction can retire only when it finishes its execution and all prior instructions have already retired. (2) Making the Speculation Primitive dependent on the execution result of the Windowing Gadget. Thus, the instructions of the Speculation Primitives cannot be executed out-of-order before the dependency is resolved. (3) Occupying the execution units or registers that are also required by the Speculation Primitive. Therefore, typical techniques used by the Windowing Gadgets include accessing non-cached memory blocks, loading memory with a chain of dependency, performing integer ALU operations with a chain of dependency~\cite{msftblogspeculative}. 

\bheading{Disclosure Gadget.} A Disclosure Gadget is a sequence of instructions that are speculatively executed, utilizing covert-channel techniques (in collaboration with the Disclosure Primitive to be explained shortly) to measure the speculation windows or the latency of data fetching, \etc.

\begin{lstlisting}[float=htbp,caption={Example of a Type-I Disclosure Gadget.},label={lst:disclosuretypeI}]
    movq (%rbx, %rcx, 1), %rbx
\end{lstlisting}

\begin{packeditemize}
\item Type-I Disclosure Gadgets only have a single memory load instruction (see \lstref{lst:disclosuretypeI}, all assembly code in this paper follows AT\&T syntax.), which we call the \textit{covert-channel sender}. A \flushreload covert channel memory buffer is allocated, which consists of 256 logically consecutive 4\kbytes pages. Each page is considered as one slot of the buffer. 
\flushreload is performed at the first integer-sized block of each page. Two forms of MOV instructions may be seen in the listings of this paper; whether or not an offset is used by MOV is determined by the values of the related registers. 


\item Type-II Disclosure Gadgets insert a sequence of ADD/SUB instructions before the covert-channel sender (see \lstref{lst:disclosuretypeII}). All these instructions have data dependencies on each other, so that they are executed in program order. The execution latency of an ADD or SUB instruction is exactly one cycle, so the total execution cycles can be estimated. An ADD and a SUB instruction are inserted in an alternating pattern so that the resulting value of their operand---the memory address used for the covert channel---does not change significantly, which simplifies the design of the covert-channel receiver. By changing the number of ADD/SUB instructions in the Windowing Gadget, the framework controls the latency of the execution of the covert-channel sender. Still, the last MOV can optionally include an offset, as shown in \lstref{lst:disclosuretypeI}.

\end{packeditemize}

\begin{lstlisting}[float=htbp,caption={Example of a Type-II Disclosure Gadget.},label={lst:disclosuretypeII}]
    [add $1, %rbx]
    [sub $1, %rbx]
    ...
    movq (%rbx), %rbx
\end{lstlisting}

Besides the three components of the instruction sequence template, the \sysname framework also incorporates a \textbf{Disclosure Primitive} to receive signals sent by the Disclosure Gadget. It leverages the \flushreload techniques to determine whether or not certain memory blocks have been accessed by the covert-channel sender of the Disclosure Gadget. As covert-channel communication is subject to noise, the test must be repeated multiple times for the Disclosure Primitive to assert whether or not the covert-channel sender was speculatively executed. \sysname only requires a binary output from the Disclosure Primitive: whether or not the signal has been received. The results will then be collected and analyzed by the Covert Channel Data Processing module.

\subsection{Speculation Primitives}
\label{subsec:instruction_model}

In this paper, we focus on Speculation Primitives that involve faults. While branch misprediction can also be explained and analyzed in the same two-phase model (see \secref{sec:spectre}), they are vulnerable by design.

To comprehensively measure all possible faults and study their exploitability, we base our tests on the exception list excerpted from the Intel Software Developer Manual \cite{IntelDevelopmentManual}. However, not all exceptions are directly related, as they do not serve the first role---they are not guarding secrets. We therefore define two templates of Speculation Primitives:

\begin{lstlisting}[float=htbp,caption={An example of a two-instruction template.},label={lst:twoinstruction}]
 // %RBX: address of a read-only page
     mov $0x42000, (%rbx)
     mov (%rbx), %rbx
\end{lstlisting}

The first template contains one single Load instruction that triggers exceptions. The second template involves two instructions, with the first causing exceptions by writing to memory or performing checks (\eg, BOUND) and the second speculatively loading data that is influenced by the first (see \lstref{lst:twoinstruction}). 
Note that the constant value 0x42000 used in the example is only for illustration purposes, which can be replaced by other values. This is also true in all the following code snippets.
If an exception is not applicable to either of these two templates, it is excluded from the analysis. We comprehensively categorize all such exceptions by their protection mechanism, with a comprehensive list given in \appref{sec:exception_list}.



\section{Understanding \vulname Vulnerabilities}
\label{sec:understanding}

\vulname vulnerabilities are caused by speculative execution. However, being able to speculatively execute instructions itself does not qualify a vulnerability. 
The root cause of  \vulname vulnerabilities is that some inaccessible secret data could be accessed by speculatively executed instructions before the processor captures the fault. Moreover, once the secret data is fetched by the speculative instructions, what can be done with it (\eg, leaking the secret using covert channels) is determined by the speculation window---the time period (in CPU cycles) of instructions executed speculatively before the faulting instruction is squashed. Our two-phase fault handling model very well separates the two race conditions:

\begin{packeditemize}
\item \textit{Race Condition I:} data fetching vs. processor fault handling.
\item \textit{Race Condition II:} covert channel transmission vs. speculative instruction squashing.
\end{packeditemize}

Race Condition I determines \textit{exploitability} and Race Condition II determines the \textit{speculation window}. The two-phase model enables a comprehensive understanding of key factors that determine these two aspects.

In this section, we \textit{first} verify the two-phase model by examining the effects of \phaseone and \phasetwo using the \sysname framework. We \textit{then} leverage \sysname to perform a systematic analysis on the two race conditions. We will show that the analysis enabled by \sysname helps us explain known attack phenomena and clarify common misunderstandings. Several tests were designed for these goals. The \sysname framework allows running each of these tests to examine different types of Speculation Primitives. For the clarification and simplicity of discussion, we illustrate these tests using a Speculation Primitive used in the Meltdown-US attack~\cite{Lipp2018meltdown}. But notice that the actual instruction sequences may differ for distinct Speculation Primitives. 









\subsection{Confirming Speculative Instruction Squash}
\label{subsec:e2_effect}

It is known that speculatively executed instructions will be squashed when the processor handles the faults. We empirically verify that \textit{issued but not yet executed \uops will not be executed after the squashing}. This fact will be the basis of the following experiments. The experiments were conducted on an Intel i7-7700HQ (KabyLake) machine with Ubuntu 16.04 (Linux 4.4.0-137) as the operating system. Each experiment is repeated for 5 times for reliability.

\begin{lstlisting}[float=htbp,caption={The effects of speculative instruction squash.}, label={lst:squash}]
 // %RBX: address of uncached covert channel buffer
 // %RDX: address of another uncached memory buffer
 // *(%RDX) = %RBX
 // %RCX: illegal address whose data is 0x42000
 // ---------------------------------------------
 // Windowing Gadget
     sub %rbx, %rcx
     movq (%rdx), %rbx
 // ---------------------------------------------
 // Speculation Primitive
     movq (%rcx, %rbx, 1), %rcx
 // ---------------------------------------------
 // Disclosure Gadget
     [add $1, %rbx]
     [sub $1, %rbx]
     ...
     movq (%rbx), %rbx
\end{lstlisting}

\bheading{Instruction sequences.}
As shown in \lstref{lst:squash}, the instructions of the Disclosure Gadget are independent of the data read by the Speculation Primitive. However, because all instructions in the Speculation Primitive and the Disclosure Gadget are dependent on the data fetched in the Windowing Gadget, the Speculation Primitive and the Disclosure Gadget start at the same time. The slow memory fetching also allows enough time for following \uops to be issued.

\bheading{Experiments and expected observations.}
In this test, the framework tunes the number of ADD/SUB instructions inserted in the Disclosure Gadget. If all issued instructions are eventually executed, we would expect to receive the covert-channel signal regardless of the number of inserted ADD/SUB instructions. Otherwise, the signal should disappear when the number of ADD/SUB instructions increases to a certain threshold.

\bheading{Results.}
In the experiments, we observed that when the inserted instructions exceed a threshold, the covert-channel receiver no longer receives any signal from the covert channel. As the number is much smaller than the size of ROB \cite{measuring-rob}, it is not caused by failed issuing due to ROB limits.

\begin{center}
\minibox[frame, rule=1pt,pad=3pt]{

\begin{minipage}[t]{0.95\columnwidth}
\textbf{Conclusion:} Issued but not yet executed \uops will be squashed when the exception is handled.
\end{minipage}
}
\end{center}


\subsection{Understanding Effects of \phaseone}

Three sets of experiments were performed to understand the effects of \phaseone on the current execution unit, other execution units, and the entire execution engine, respectively. The experiments were performed in the same settings as \secref{subsec:e2_effect}.


\subsubsection{\phaseone on Current Execution Unit}
\label{subsec:e1_existence}

This test is designed to determine how \phaseone affects the execution unit being used by the Speculation Primitive. 

\begin{lstlisting}[float=htbp,caption={The effects of \phaseone on the current execution unit.},label={lst:phaseonecurrent}]
 // %RBX: address of uncached covert channel buffer
 // %RDX: address of another uncached memory buffer
 // *(%RDX) = %RBX
 // %RCX: illegal address whose data is 0x42000
 // ---------------------------------------------
 // Windowing Gadget 
     movq (%rdx), %rdx
 // ---------------------------------------------
 // Speculation Primitive
     movq (%rcx), %rcx
 // ---------------------------------------------
 // Disclosure Gadget
     movq (%rbx, %rcx, 1), %rbx
\end{lstlisting}

\bheading{Instruction sequences.}
As shown in \lstref{lst:phaseonecurrent}, the instruction sequence consists of a Windowing gadget, a Speculation Primitive, and a Disclosure Primitive. The Speculation Primitive is a simple slow memory load to ensure that the retirement latency of the Speculation Primitive remains constant by postponing it to a late enough fixed time. Thus, influence of \phasetwo is excluded from this experiment. 

\bheading{Experiments and expected observations.}
The tests were repeated four times, with the secret data placed in L1D cache, L2 cache, LLC, and memory, respectively. The TLB entry of the secret's address is always flushed to ensure a fixed \phaseone latency. In these experiments, we would hope to see whether the changes of the data fetching latency affects the covert-channel signal received by the Disclosure Primitive. If so, whether \phaseone happens before the data is fetched affects the return values of current execution unit.

\bheading{Results.}
We observed that only when the secret data is stored in the L1 cache could the correct covert-channel signal be received. When the secret data is store in L2, LLC, or the memory, a zero signal is received. This observation validates our theory in \ref{subsec:handling}. The execution unit terminates after catching the exception; a dummy value of zero is returned as the result of the execution. 
We will show that \phasetwo is not relevant in this experiment in \secref{subsec:e2_change}.

\begin{center}
\minibox[frame, rule=1pt,pad=3pt]{

\begin{minipage}[t]{0.95\columnwidth}
\textbf{Conclusion:} {\phaseone} terminates the current execution unit. If the latency of {\phaseone} is greater than the data fetching latency, the correct value can be propagated to the speculative instructions; otherwise, a zero value will be returned.
\end{minipage}
}
\end{center}




\subsubsection{\phaseone on Other Execution Units}
\label{subsec:e1_effect}

As \phaseone terminates the current execution unit, it is directly related to the exploitability. However, it is not yet clear the exploitability is also affected by \phaseone of other faulting instructions.
If so, attacks may be enhanced by combining two or more Speculation Primitives.

\bheading{Instruction sequences.}
Different from other tests, two Speculation Primitives are included in this test to determine whether \phaseone of the first Speculation Primitive also influences the execution unit used by the second Speculation Primitive. We consider two cases, depending on whether the two Speculation Primitives access the same memory address.

First, the two Speculation Primitives access the same memory address. The instruction sequence is designed as shown in \lstref{lst:sameaddress}. A sequence of inter-dependent ADD/SUB instructions are inserted between the two Speculation Primitives to control the delay of the execution of the second Speculation Primitive. Since the first inserted ADD instruction and the first Speculation Primitive both have data dependency on the Windowing Gadget, they start at the same time. But the second Speculation Primitive has to wait until all the inserted ADD/SUB instructions finish. The Disclosure Gadget is used to monitor the data fetched by the second Speculation Primitive.

\begin{figure}[tb]
\centering
\includegraphics[width=\columnwidth]{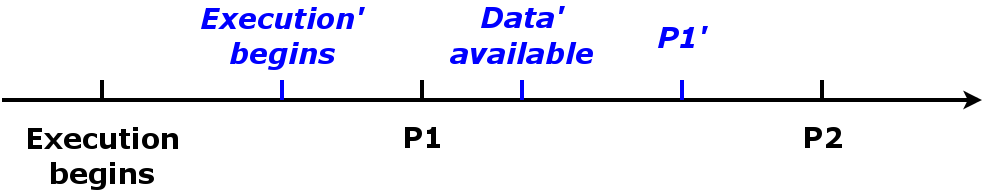}
\caption{Illustration of \phaseone effects on other execution units.}
\label{fig:phaseone_on_other}
\end{figure}

Second, the two Speculation Primitives access different memory addresses. The instruction sequence used is the same as the first case, except that the memory addresses accessed by the two Speculation Primitives are different.

\begin{lstlisting}[float=htbp,caption={The effects of P1 on the other execution units},label={lst:sameaddress}]
 // %RBX: address of uncached covert channel buffer
 // %RDX: address of another uncached memory buffer
 // *(%RDX) = %RBX
 // %RCX: illegal address #1
 // %RAX: illegal address #2 whose data is 0x42000
 // ---------------------------------------------
 // Windowing Gadget
     sub %rbx, %rax
     sub %rbx, %rcx
     movq (%rdx), %rbx
 // ---------------------------------------------
 // Speculation Primitive #1
     movq (%rcx, %rbx, 1), %rcx
 // ---------------------------------------------
 // special inserted instructions for the experiment
     add %rbx, %rax
     [add $1, %rax]
     [sub $1, %rax]
     ...
 // ---------------------------------------------
 // Speculation Primitive #2
     movq (%rax), %rax
 // ---------------------------------------------
 // Disclosure Gadget
     movq (%rbx, %rax, 1), %rbx
\end{lstlisting}

\bheading{Experiments and expected observations.} In both tests, the relevant data are stored in the L1D cache. As shown in \figref{fig:phaseone_on_other}, the events with black labels describe the first Speculation Primitive, and the ones with blue labels describe the second. By delaying the execution of the second Speculation Primitive, \phaseone of the first Speculation Primitive can happen before the second Speculation Primitive fetches the secret data. Therefore, in the experiment, we gradually inserted more instructions between the two Speculation Primitive to delay the start of the second Speculation Primitive. 

\bheading{Results.}
In both experiments, regardless of whether the two Speculation Primitives accesses the same memory addresses or not, by gradually inserting instructions between them, we never witnessed that the received covert-channel signal changes from the correct value to zero. Instead, we only observed that after some threshold, the signal disappears. This suggests that \phaseone of the first Speculation Primitive does not influence the other execution units (by zeroing their results) but its \phasetwo does (by squashing their execution).
Therefore, {\phaseone} of the first Speculation Primitive does not affect other execution units.

\begin{center}
\minibox[frame, rule=1pt,pad=3pt]{

\begin{minipage}[t]{0.95\columnwidth}
\textbf{Conclusion:} Performing transient execution attacks with two or more Speculation Primitive does not increase the exploitability.
\end{minipage}
}
\end{center}

\subsubsection{\phaseone on Execution Engine}
\label{subsec:e1_exec}

To confirm that \phaseone has nothing to do with the speculation window, we show that \phaseone does not influence other components of the execution engine, \eg, by squashing speculative \uops in ROB or altering code fetching in the front end. 

Particularly, we define \textit{speculation window} as the maximal number of CPU cycles from the beginning of the speculative execution till all speculatively executed instructions are squashed. 
\sysname enables us to indirectly measure the speculation window in the following test. 

\bheading{Instruction sequences.}
The design is close to \lstref{lst:squash}. The only difference is that after Line 9, a memory load instruction (\ie, movq (\%rax, \%rbx, 1), \%rax) is added. 

\bheading{Experiments and expected observations.}
The strategy of the test is to change the latency of \phaseone while fixing the latency of \phasetwo. If the measured speculation window does not change according to \phaseone latency, \phaseone has no effect on the entire execution engine.

To change the latency of \phaseone, we control the TLB status of the page storing the secret data---by preloading or flushing the TLB entry. In this way, \phaseone of the Speculation Primitive changes accordingly. 

To fix \phasetwo latency, one additional memory load instruction is added in the windowing gadget, which begins to execute at the same time as the Speculation Primitive and the Disclosure Gadget. The goal of this instruction is to delay the retirement of all subsequent instructions, so that the retirement of the Speculation Primitive waits on the retirement of this memory load instruction. In this way, the \phasetwo latency is not determined by the the Speculation Primitive itself, which changes according to TLB presence, but by the retirement of the memory load instruction. To achieve this goal, the data to be loaded by this instruction is placed in L2 cache and the TLB of the corresponding page is flushed.

\bheading{Results.}
When changing the \phaseone latency, we did not observe any changes in the speculation window by counting the maximal ADD/SUB instruction numbers in the Disclosure Gadget that still allows the last covert channel access instruction to execute. 

\begin{center}
\minibox[frame, rule=1pt,pad=3pt]{

\begin{minipage}[t]{0.95\columnwidth}
\textbf{Conclusion:} \phaseone does not affect the entire execution engine; altering \phaseone does not change the speculation window. 
\end{minipage}
}
\end{center}

\subsection{Understanding Effects of \phasetwo}
\label{subsec:e2_change}


The following test aims to confirm that \phasetwo squashes all speculative instructions and \phasetwo can be manipulated.

\begin{lstlisting}[float=htbp,caption={Tuning \phasetwo latency.},label={lst:retirement}]
 // %RBX: address of uncached covert channel buffer
 // %RCX: illegal address whose data is 0x42000
 // ---------------------------------------------
 // Windowing Gadget
     movapd \%xmm0, \%xmm1
     addpd \%xmm1, \%xmm0
     [cpuid]
     mulpd \%xmm1, \%xmm0
     ...
     movapd \%xmm0, \%xmm1
     addpd \%xmm1, \%xmm0
     mulpd \%xmm1, \%xmm0
 // ---------------------------------------------
 // Speculation Primitive
     movq (%rcx), %rcx
 // ---------------------------------------------
 // Disclosure Gadget
     [add $1, %rcx]
     [sub $1, %rcx]
     ...
     movq (%rbx, %rcx, 1), %rbx
\end{lstlisting}

\bheading{Instruction sequences.}
The instruction sequence is shown in \lstref{lst:retirement}.
As the retirement of the Speculation Primitive only happens after that of the Window Gadget, the strategy is to manipulate the latter and look for any changes in the spelucation window.. The Windowing Gadget consists of 25 repeated sequences of three SSE2 instructions: MOVAPD, ADDPD and MULPD---thus 75 instructions in total. These floating point instructions are slow but can be executed in parallel with the Speculation Primitive and the Disclosure Gadget. Each of these floating point instructions has data dependency on its predecessor. 

To fine tune the retirement of the Speculation Primitive, a CPUID instruction is inserted in the Windowing Gadget. As the CPUID instruction serializes the execution of the instructions before and after it (\ie, no instruction is issued before CPUID retires), only the floating point instructions after CPUID are effective in the Windowing Gadget. Therefore, the retirement of the Speculation Primitive is further delayed if there is CPUID is inserted earlier in the floating point instructions; and vice versa.
The Disclosure Gadget is of type-II. All instructions are dependent on the Speculation Primitive.



\bheading{Experiments and expected observations.}
For each position 
\begin{wrapfigure}{r}{0.55\columnwidth}
\includegraphics[width=\linewidth]{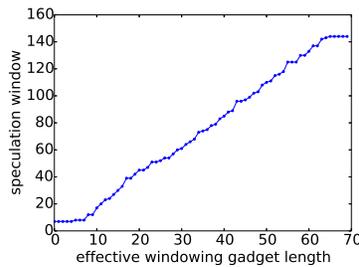}
    \caption{Size of the effective Windowing Gadget vs. speculation windows.}
    \label{fig:atk_window_xmm}
\end{wrapfigure}
in the Windowing Gadget where CPUID is inserted, the framework automatically alter the number of ADD/SUB instructions to identify the speculation window as in \secref{subsec:e1_exec}. In this way, the correlation between \phasetwo and the speculation window can be observed. 

\bheading{Results.}
In \figref{fig:atk_window_xmm}, the x-axis is the effective size of Windowing Gadget (tuned by moving the position of CPUID) and the y-axis is
the speculation window. By gradually moving the position of CPUID from the beginning of the Windowing Gadget to the end, which changes the effective length of Windowing Gadget, the speculation window also grows accordingly. 
Prior studies~\cite{Lipp2018meltdown} have reported the speculation window of certain attack variants. However, our experiment suggests that it is not meaningful to report the size of the speculation window as it can be changed in manners described above. Despite that, the speculation window is still limited by the size of the ROB. 
In \figref{fig:atk_window_xmm}, the maximum speculation window is about 140 cycles, reflecting the ROB size of greater than 140 instructions (as each ADD/SUB instruction takes 1 cycle). And this is already big enough for covert channels to transmit data through one memory operation at a time.

\begin{center}
\minibox[frame, rule=1pt,pad=3pt]{

\begin{minipage}[t]{0.95\columnwidth}
\textbf{Conclusion:} The speculation window of any Speculation Primitive can be altered by delaying its \phasetwo. 
\end{minipage}
}
\end{center}

\subsection{Investigating Race Conditions}
\label{subsec:race}

A successful exploitation of \vulname depends on the outcome of the race conditions: (i) Data fetching latency must be lower than \phaseone latency, and (ii) speculative covert channel transmission should be faster than \phasetwo. As such, we leverage \sysname to investigate the following questions: 

\begin{packeditemize}

\item Is it possible to quantitatively measure the race conditions?

\item Is it possible to control the outcomes of the race conditions?

\end{packeditemize}


\subsubsection{Revisiting Race Condition II}


\secref{subsec:e2_change} already demonstrated that we are able to quantitatively measure Race Condition II by evaluating the speculation window. 
Moreover, \figref{fig:atk_window_xmm} suggests that by altering the retirement of the Windowing Gadget, the attacker is able to delay \phasetwo of the Speculation Primitive. 

\begin{center}
\minibox[frame, rule=1pt,pad=3pt]{

\begin{minipage}[t]{0.95\columnwidth}
\textbf{Conclusion:} The attacker can always win Race Condition II by delaying \phasetwo of the Speculation Primitive. 
\end{minipage}
}
\end{center}



\subsubsection{Measuring \vulname Exploitability}
\label{subsec:e1_measure}

As the outcome of Race Condition II can be controlled, a successful attack depends solely on the outcome of Race Condition I. Therefore, by measuring the outcome of Race Condition I, \sysname enables automated tests of the \textit{exploitability} of all possible exception-based variants on various processors. It leverages its ability to enumerate possible combinations of execution conditions such as cache and TLB presence to determine the exploitability under the most optimal condition.

We performed tests on 9 machines (2 laptops, 5 desktops and 1 cloud VM). All tested machines run Ubuntu 16.04 with Linux kernel 4.4.0-137 (or 4.4.0-141 for compatibility issues on newer hardware) and KPTI is turned off. 
All the Intel microcode versions are rolled back to version 20171117 (except for Coffee Lake, which does not have older microcode, and the cloud VM, which we cannot control). 
The AMD microcode version is 3.20180515.1. The test can be extended to evaluating patched microcode and countermeasures, such as KPTI~\cite{linuxpatch} and PTE inversion~\cite{pteinversion}. We would like to open source \sysname to enable other researchers to perform tests in other settings.


\bheading{Instruction sequences.}
As shown in \lstref{lst:exploitability}, the Windowing Gadget has a memory load instruction that retires slowly due to long latency of memory access, but it does not have dependency on the previous instructions, nor does any subsequent instructions depend on it. It is used to ensure that the retirement of the Speculation Primitive (the \phasetwo latency) is sufficiently delayed. 

\begin{lstlisting}[float=htbp,caption={Exploitability test with \phaseone measurement.},label={lst:exploitability}]
 // %RBX: address of uncached covert channel buffer
 // %RDX: address of another uncached memory buffer
 // *(%RDX) = %RBX
 // %RCX: illegal address whose data is 0x42000
 // ---------------------------------------------
 // windowing gadget
     movq (%rdx), %rdx
 // ---------------------------------------------
 // speculation primitive
     movq (%rcx), %rcx // could be any illegal inst.
 // ---------------------------------------------
 // disclosure gadget
     movq (%rbx, %rcx, 1), %rbx
\end{lstlisting}

\bheading{Experiments and expected observations.}
Each tested instruction sequence was executed under a variety of conditions, with varying data access latency (cached in L1D, L2, or LLC) and address translation latency (whether or not TLB entries are created for the corresponding pages). \sysname is able to enumerate all possible combinations to achieve the optimal condition. In each test, if the Disclosure Primitive receives the correct signal from the covert channel, the vulnerability is exploitable. Otherwise, if a zero signal is received, the vulnerability is not exploitable as the \phaseone latency is shorter than data available latency. However, if no signal can be received, it suggests speculation is not allowed by the variant.


\begin{table*}[tb]
\begin{center}
{\scriptsize
\begin{tabular}{p{0.34\columnwidth}|p{0.13\columnwidth}|p{0.13\columnwidth}| p{0.14\columnwidth}|p{0.14\columnwidth}|p{0.16\columnwidth}|p{0.13\columnwidth}|p{0.12\columnwidth}|p{0.15\columnwidth}|p{0.13\columnwidth}}
 & Laptop 1 & Laptop 2 & Desktop 1 & 
Desktop 2 & Desktop 3 & Desktop 4 & Desktop 5 & Desktop 6 & Cloud 1 \\ 
Variant & KabyLake & KabyLake & Haswell-EP & SandyBridge & Westmere-EP & CoffeeLake & KabyLake & AMD EPYC & Skylake-SP \\
\toprule

PTE (Present) & Y & N/A & N/A & N/A & N/A & Y & Y & N/A & Y \\ \hline

PTE (Reserved) & Y & N/A & N/A & N/A & N/A & Y & Y & N/A & Y \\ \hline

PTE (US) & Y & Y & Y & Y & Y & Y & Y & R & Y \\ \hline

Load CR4 & R & R & R & R & R & R & R & R & R \\ \hline

Load MSR (0x1a2) & R & R & R & R & R & R & R & N/A & N/A \\ \hline

Protection Key (User) & N/A & N/A & N/A & N/A & N/A & N/A & N/A & N/A & Y \\ \hline

Protection Key (Kernel) & N/A & N/A & N/A & N/A & N/A & N/A & N/A & N/A & Y \\ \hline

SMAP violation & Y & Y & N/A & N/A & N/A & Y & Y & Y* & Y** \\ \hline

PTE (write w/ RW=0) & Y & Y* & Y & Y & Y & Y & Y & R & Y \\ \hline

Load xmm0 (CR0.TS) & Y & Y & Y & Y & Y & Y & Y & N/A & N/A \\ \hline

BOUND (32-bit) & Y & Y & Y & Y & Y & Y & Y & Y & Y \\ \hline

DS Over-Limit (32-bit) & N & N & N & N & N & N & N & Y & N \\ \hline

SS Over-Limit (32-bit) & N & N & N & N & N & N & N & Y & N/A \\ \hline

DS Not-Present (32-bit) & R & R & R & R & R & R & R & R & R \\ \hline

SS Not-Present (32-bit) & R & R & R & R & R & R & R & R & R \\ \hline

DS Execute-Only (32-bit) & R & R & R & R & R & R & R & R & R \\ \hline

CS Execute-Only (32-bit) & R & R & R & R & R & R & R & R & R \\ \hline

DS Read-Only (write, 32-bit) & Y & Y & Y & Y & Y & Y & Y & R & Y \\ \hline

SS Read-Only (32-bit) & R & R & R & R & R & R & R & R & R \\ \hline

DS Null (32-bit) & N & N & N & N & N & N & N & R & N \\ \hline

SS Null (32-bit) & R & R & R & R & R & R & R & R & R \\ \hline

SS $DPL \neq CPL$ (32-bit) & R & R & R & R & R & R & R & R & R

\end{tabular}
}
\end{center}
\vspace{-10pt}
\caption{Exploitability evaluation on different machines. Y: exploitable. N: non-exploitable. R: no speculative execution. N/A: unable to test. Y*: both expected data and zero data are captured in all covert channel reloads. Y**: the only exploited case with success rate not 100\% (about 60\%). Laptop 1: i7-7820HQ (Kaby Lake). Laptop 2: i7-7700HQ (Kaby Lake). Desktop 1: Xeon E5-1607v3 (Haswell-EP). Desktop 2: i3-2120 (Sandy Bridge). Desktop 3: Xeon E5620 (Westmere-EP). Desktop 4: Xeon E-2124G (Coffee Lake). Desktop 5: i7-7700 (Kaby Lake). Cloud 1: Amazon EC2 C5 instance. Desktop 6: AMD EPYC 7251.}
\label{tbl:exploitability}
\end{table*}

\bheading{Results.}
The results are shown in \tabref{tbl:exploitability} and a reference of the tested variant names to their description can be found in \appref{sec:exception_list}. Some variants are unable to be tested due to lack of hardware support or OS support on certain machines and they are marked as N/A. On the tested Intel machines, Meltdown-US (accessing supervisor memory page from user space) and Meltdown-RW (writing to a read-only memory page) are exploitable while the AMD machine shows no speculation. When testing Meltdown-Present (\texttt{Present} flag cleared) and Meltdown-Reserved (\texttt{Reserved} flag set), signal handler cannot be used since the whole OS will crash. Thus, only the machines also equipped with Intel TSX~\cite{inteltsx} are tested and reported. Loading restricted registers (CR4 and MSR) on all tested machines show that no speculative load is allowed. Meltdown-MPK (bypassing the restriction of memory protection keys) could only be tested on Amazon EC2 E5 instance and it was found exploitable. Meltdown-FP (accessing a lazy-context-switch float pointer register) is also found exploitable on Intel machines. Meltdown-BR ( accessing an array with an over-range index) is found exploitable on all tested machines although the BOUND instruction raises an exception when it discovers the violation. In 32-bit mode, segmentation is enabled and thus relevant variants could be tested. Most of them are not exploitable or does not allow further speculation at all, since paging checks has to wait until segmentation translation produces a linear address while segmentation check begins at the same time as the translation. However, still some of them are found exploitable on either Intel or AMD platform.



In our experiments, a few new variants were found by the \sysname framework. \textit{We have reported these new variants to Intel and AMD already.}

\begin{packeditemize}
\item \textit{Supervisor mode access violating SMAP (Intel \& AMD).}
Supervisor Mode Access Prevention (SMAP) forbids code in the kernel mode from accessing to user-space addresses.  However, as shown in \tabref{tbl:exploitability}, on some of our tested machines, SMAP can be bypassed using speculative execution when the secret data is in the L1 cache. When the data is in lower-level caches, zero signal is captured. 

\item \textit{Supervisor mode access bypassing MPK (Intel).}
When a user space page is set to be inaccessible using Memory Protection Key (MPK), its accesses from kernel code is also forbidden, which triggers a page fault exception. However, as demonstrated on the tested cloud server (see \tabref{tbl:exploitability}), if the secret data is already in the L1 cache, it can be leaked through speculatively executed Disclosure Gadget. 

\item \textit{Memory writes to read-only data segments (Intel).}
Segmentation is used in the 32-bit mode. However, memory writes to a read-only data segments can be speculatively used by following instructions. In this case, the caching and TLB status does not affect the exploitability. This vulnerability is similar to the \texttt{Meltdown-RW} \cite{kiriansky2018speculative}, but its security implication is different: \texttt{Meltdown-RW} takes advantage of store-to-load forwarding. As the store buffer is indexed by the linear address (not logical address), the forwarding is speculative as the logical-to-physical translation is not yet finished. Therefore, it is not surprising that the permission check (using flags in the PTE) happens after the store-to-load forwarding. However, in contrast, the segmentation check is performed during the translation from logical addresses to linear addresses. Thus, when the store buffer queues the store instruction to the given linear address \cite{US6378062}, the segmentation access privilege check should already be done. However, our test suggests that this is not the case as a following load could directly use the store value. This validates the conclusion of the recent Fallout attack \cite{fallout} that store buffers predict aggressively using only the lowest bits of addresses.

\item \textit{Reading from a logical address over the limit of segment (AMD).}
When tested in 32-bit mode, a load to a logical address beyond the segment limit is forbidden, which triggers an exception. However, we have found that the segmentation check can be bypassed by speculative execution. The vulnerability is exploitable when the TLB entry of the page is present and the data is cached in the L1 cache. The same vulnerability cannot be found in Intel processors.

\end{packeditemize}



\begin{center}
\minibox[frame, rule=1pt,pad=3pt]{

\begin{minipage}[t]{0.95\columnwidth}
\textbf{Conclusion:} \sysname enables automated tests of \vulname vulnerabilities on various processors. It detects several new variants of transient execution attacks.
\end{minipage}
}
\end{center}

\bheading{Extended study on state-of-the-art mitigation.}
KPTI~\cite{linuxpatch} is a software solution designed to prevent the exploitation of the Meltdown-US vulnerability, but not removing the vulnerability from hardware. Since the test cases of \sysname are designed by explicitly setting the bits in PTE, KPTI should not place any influence on it. On the other hand, the existing microcode patches are for Spectre variants and L1TF only~\cite{microcodesummary} and thus is expected not to affect the tests. We performed the tests on Laptop 1 with the latest microcode patch and found all vulnerabilities still present, including Meltdown-US.

\subsubsection{Quantitatively Measuring \phaseone Latency}
\label{subsec:e1_change}

A more powerful measurement could be done with \sysname to quantitatively measure the relative latency of \phaseone compared to data fetching. 

\begin{lstlisting}[float=htbp,caption={Quantitative measurement of \phaseone latency},captionpos=b,label={lst:framework2}]
 // %RBX: address of uncached covert channel buffer
 // %RCX: illegal address whose data is 0x42000
 // ---------------------------------------------
 // Suppressing Primitive
     [movq (%rax), %rax] // legal access
     [movq (%rax), %rax] // legal access
     ...
     movq (%rax), %rax // suppressing w/ exception
 // ---------------------------------------------
 // Speculation Primitive
     movq (%rcx), %rcx // could be any illegal inst.
 // ---------------------------------------------
 // Disclosure Gadget
     [add $1, %rcx]
     [sub $1, %rcx]
     ...
     movq (%rbx, %rcx, 1), %rcx
\end{lstlisting}

\bheading{Instruction sequences.}
The construction of the instruction sequence is different from other tests as a \textit{Suppressing Primitive} is needed in the test. The Suppressing Primitive precedes all other components in order to conceal the effect of executing the \pluginonename. This is achieved by ensuring that the instructions executed in all other components never retire. In the example of \lstref{lst:framework2}, the Suppressing Primitive is simply an illegal memory load from address 0, but it can also be implemented using conditional branches, indirect jumps, or retpoline. The Suppressing Primitive creates a fixed speculation window for the rest of the instruction sequence to execute. Because it is desired that this speculation window is greater than the one created by the Speculation Primitive, the Suppressing Primitive includes a few memory loads before the faulting instruction and leverages the pointer chasing technique~\cite{pointerchasing} to further enlarge its speculation window. The speculation window cannot be longer than the size of the ROB.

A type-II Disclosure Gadget is used; all its instructions depend on the Speculation Primitive. Thus, the Disclosure Gadget only begins execution after the data is returned from the Speculation Primitive, regardless of its correctness.

\begin{figure}[t]
\centering
\includegraphics[width=\columnwidth]{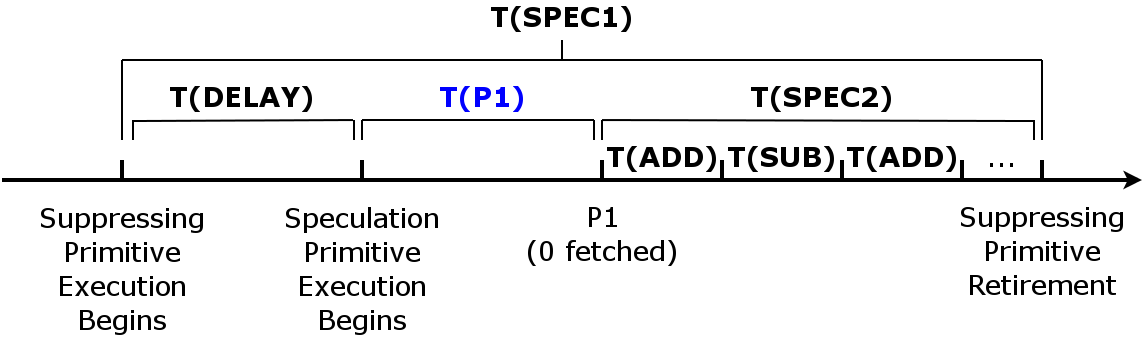}
\caption{Illustration of quantitatively measurement of \phaseone latency.}
\label{fig:e1_latency}
\vspace{5pt}
\end{figure}

\bheading{Experiments and expected observations.}
By flushing the secret data to memory, the framework ensures that a zero data is returned by the Speculation Primitive. Let $T_{P1}$ be the number of cycles to retrieve the zero data by the Speculation Primitive, $T_{spec1}$ be the speculation window of the Suppressing Primitive, and $T_{delay}$ be the latency for the Speculation Primitive to begin execution, which are both fixed. The number of inserted instructions in the Disclosure Gadget is tuned to determine $T_{spec2}$, which is the speculation window of the Speculation Primitive. As the Disclosure Gadget starts after the Speculation Primitive has retrieved the data, we have $T_{delay} + T_{P1} + T_{spec2} = T_{spec1}$ (shown in \figref{fig:e1_latency}).

To estimate $T_{P1}$, a control test is run. In the control test, the Speculation Primitive accesses a legal data, which does not trigger exceptions. But still, the execution will be reverted due to the Suppressing Primitive. In this case, the number of cycles to retrieve the legal data is $T_{data}$. With all other conditions unchanged, we have $T_{delay} + T_{data} + T_{spec2'} = T_{spec1}$. Without the need of calculating the exact values of $T_{spec1}$ and $T_{delay}$, the relationship between $T_{data}$ and $T_{P1}$ can be estimated in a differential manner: $T_{data} - T_{P1} = T_{spec2} - T_{spec2'}$.

\bheading{Results.}
We ran the tests on Meltdown-US as an example and found that $T_{data} - T_{P1} = 0$, which means \phaseone and L1 cache data fetching arrives at the same cycle. This also explains why most attacks cannot work when the data is in higher-level caches (unless the effects of prefetching is exploited, see \secref{subsec:prefetch}).  


We also performed the tests on negative results in \tabref{tbl:exploitability}. For example, when accessing data beyond the segment limit, data in \tabref{tbl:exploitability} suggest the vulnerability is not exploitable. By running the test to quantitatively measure $T_{P1}$, we found out that $T_{data} - T_{P1} = 0$, which means \phaseone come right before the data is available.  
However, when data is available in the L2 cache, we find $T_{data} - T_{P1} = -12$, which suggests \phaseone come 12 cycles earlier than the data is available. Due to \sysname's ability to quantitatively measure a negative relative \phaseone latency, we are able to affirmatively claim the inexploitability of certain variants on given hardware. 


\begin{center}
\minibox[frame, rule=1pt,pad=3pt]{

\begin{minipage}[t]{0.95\columnwidth}
\textbf{Conclusion:} \sysname enables quantitative measurement of Race Condition I, providing security assurance for the negative results of the exploitability tests.  
\end{minipage}
}
\end{center}

\subsubsection{Controlling Race Condition I}
\label{subsubsec:e1_control}

The ability to quantitatively measure the relative \phaseone latency also enables the exploration of the controllability of Race Condition I. Given a certain variant of \vulname vulnerabilities, we can leverage \sysname to alter one factor (\eg, TLB entry status) while keeping all others unchanged. Then, \sysname is able to determine how the relative \phaseone changes according to the tested factor. For example, we found that the absence of TLB entry leads to a decrease of the relative \phaseone latency by over 100 cycles when testing the variant violating segmentation limit. However, we did not find such an effect when testing it on Meltdown-US attacks. We leave a comprehensive examination of all variants and all possible factors to future work. 

\subsection{Speculation Primitive as Prefetcher}
\label{subsec:prefetch}

Given the study on the race conditions for exploitation in ONE round of attack, an attack could be guided to create an optimal attack scenario. However, not all of the resources are within the control of the attacker. Thus, the following question is whether the attacker is able to change those conditions that controls/influences the race condition. How a sophisticated attacker can manipulate victim is beyond the scope of this paper, so we only study whether ONE round of attack itself benefits the race conditions of next round of attack. Take Meltdown-US and Meltdown-P (the base of L1TF) as an example, the only influencing condition towards the exploitability is data fetching latency as analyzed in \secref{subsubsec:e1_control}. We are unable to test buffers for now, so we focus on caching.

Some Speculation Primitive may only lead to exploitable vulnerability when the data fetching latency is small---the data is already cached in the L1 cache---but others may succeed even when the data is completely uncached. We speculate the root cause is that some Speculation Primitive, though failed to extract secret from L2, LLC, or memory, could work as a prefetcher to preload secret data into L1 caches so as to facilitate future attacks of the same kind. We empirically validate this hypothesis.

\bheading{Experiments.}
In this test, Windowing Gadget and Disclosure Gadget are unnecessary. The experiment is conducted in three steps: First, the secret data to be accessed by the Speculation Primitive is preloaded (from the shadow virtual address) into the LLC or memory. Second, the tested Speculation Primitive is executed $N$ times (with the exceptions suppressed by the \sysname framework). Third, the data is reloaded from the shadow address and the latency is measured. Two versions of this experiment were tested: (1) $N=0$; (2) $N=1000$.

\begin{figure}[t]
\centering
\subfloat[Data in LLC, w/o prefetching.]{
    \includegraphics[clip,width=0.47\columnwidth]{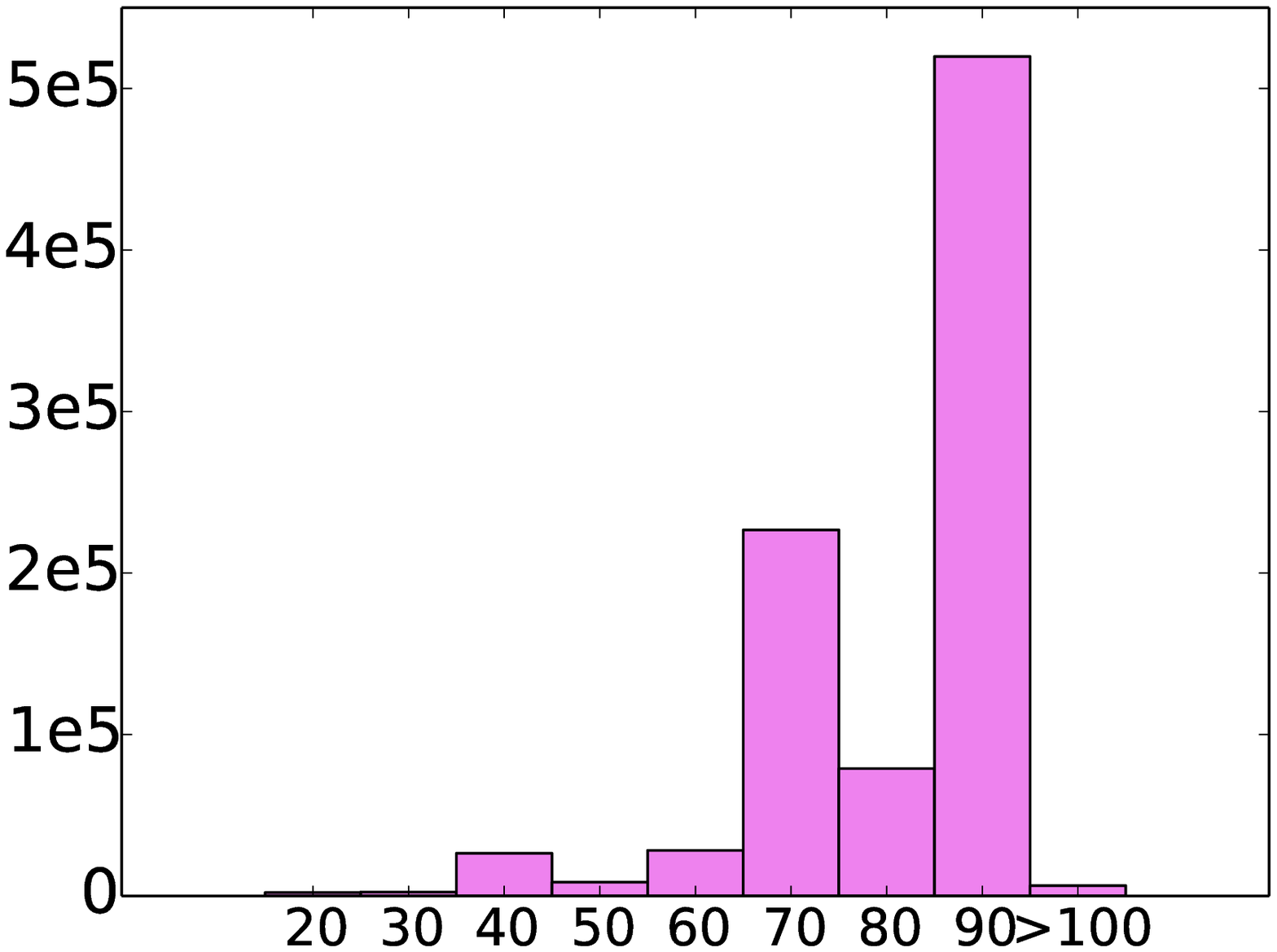}
    \label{fig:prefetch_l3_standard}
}
\subfloat[Data in LLC, w/ prefetching.]{
    \includegraphics[clip,width=0.47\columnwidth]{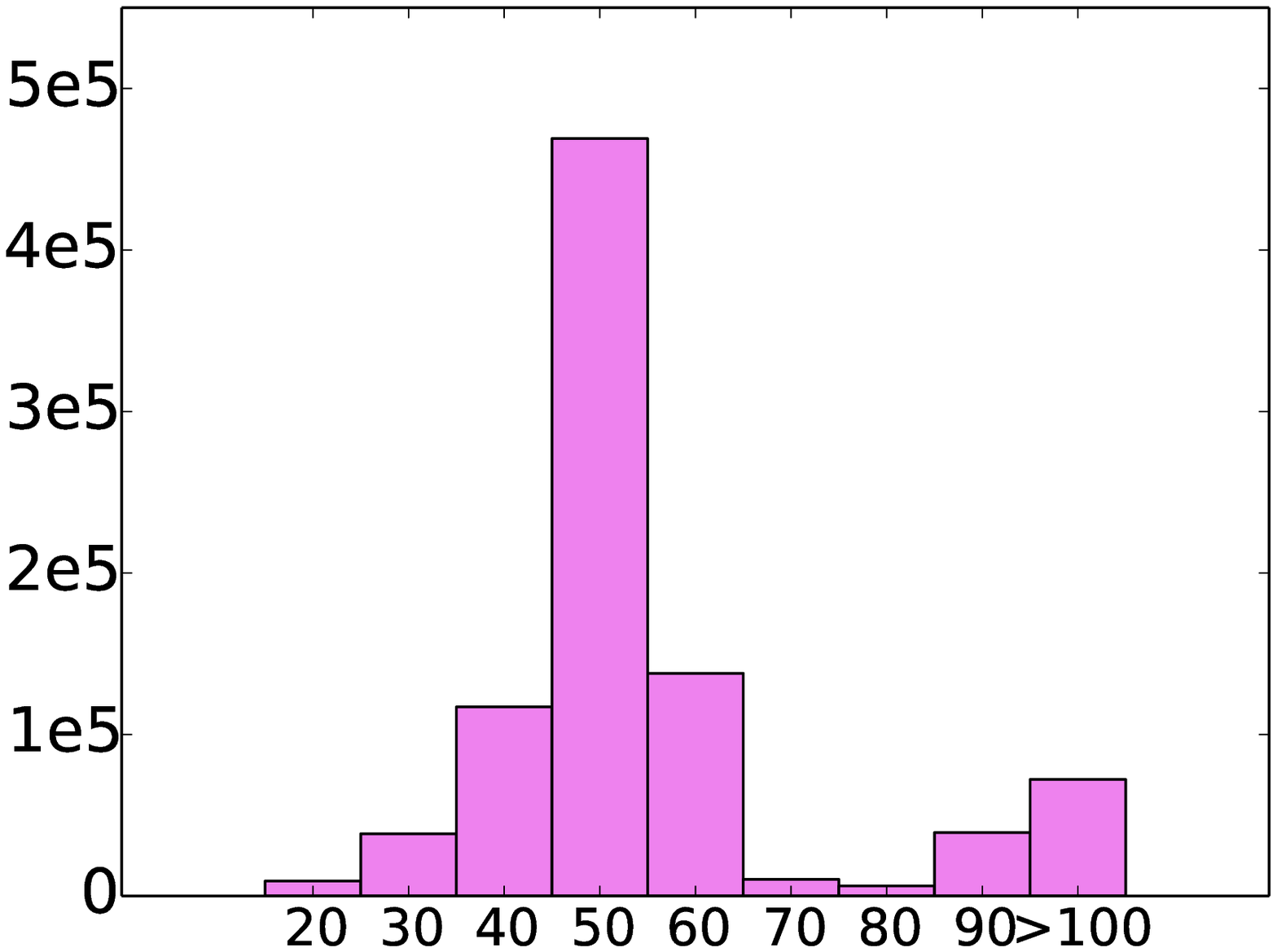}
    \label{fig:prefetch_l3_1000}
}

\subfloat[Data in memory, w/o prefetching.]{
    \includegraphics[clip,width=0.47\columnwidth]{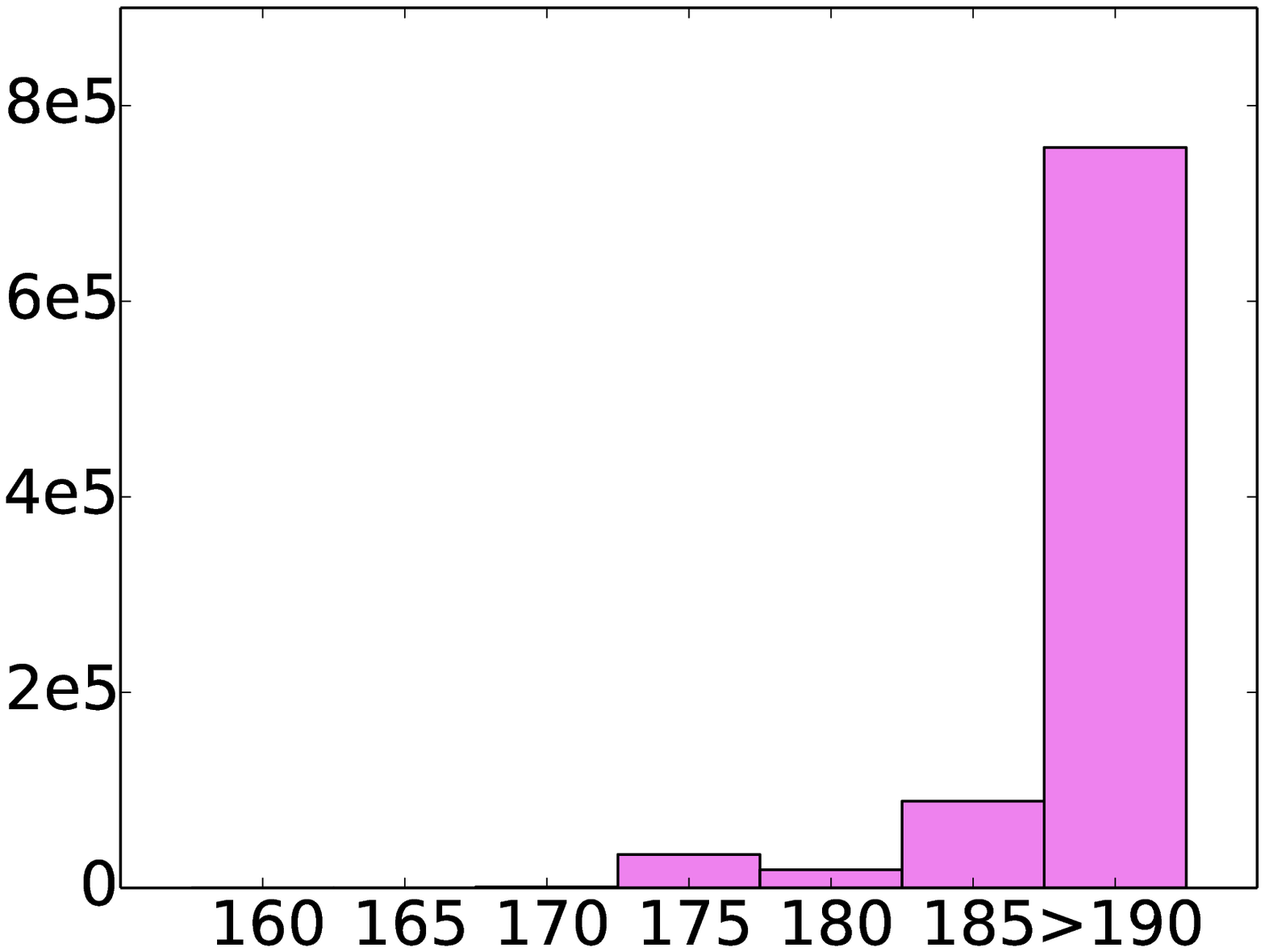}
    \label{fig:prefetch_memory_standard}
}
\subfloat[Data in memory, w/ prefetching.]{
    \includegraphics[clip,width=0.47\columnwidth]{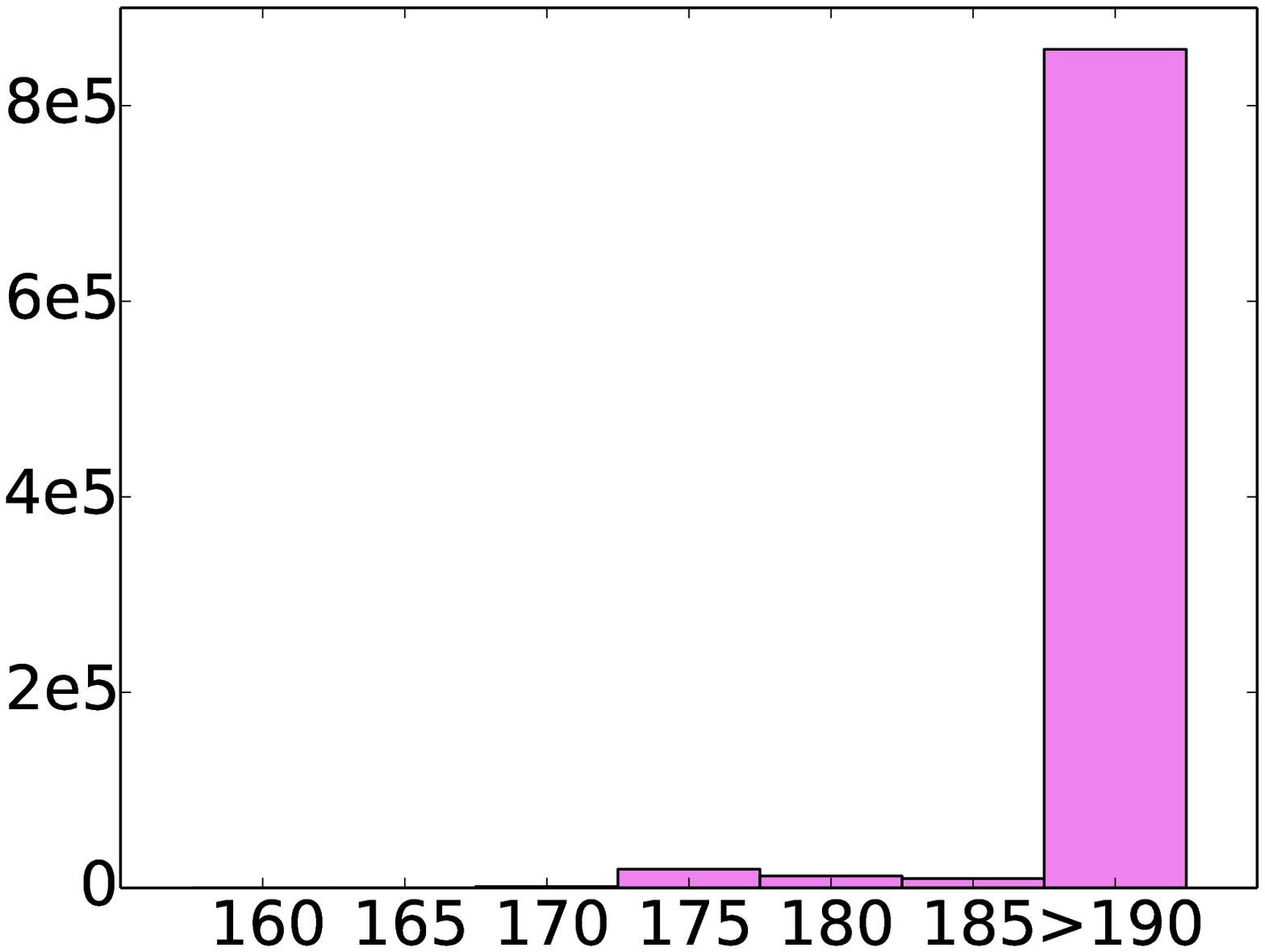}
    \label{fig:prefetch_memory_1000}
}

\subfloat[Terminal fault, data in LLC, w/o prefetching.]{
    \includegraphics[clip,width=0.47\columnwidth]{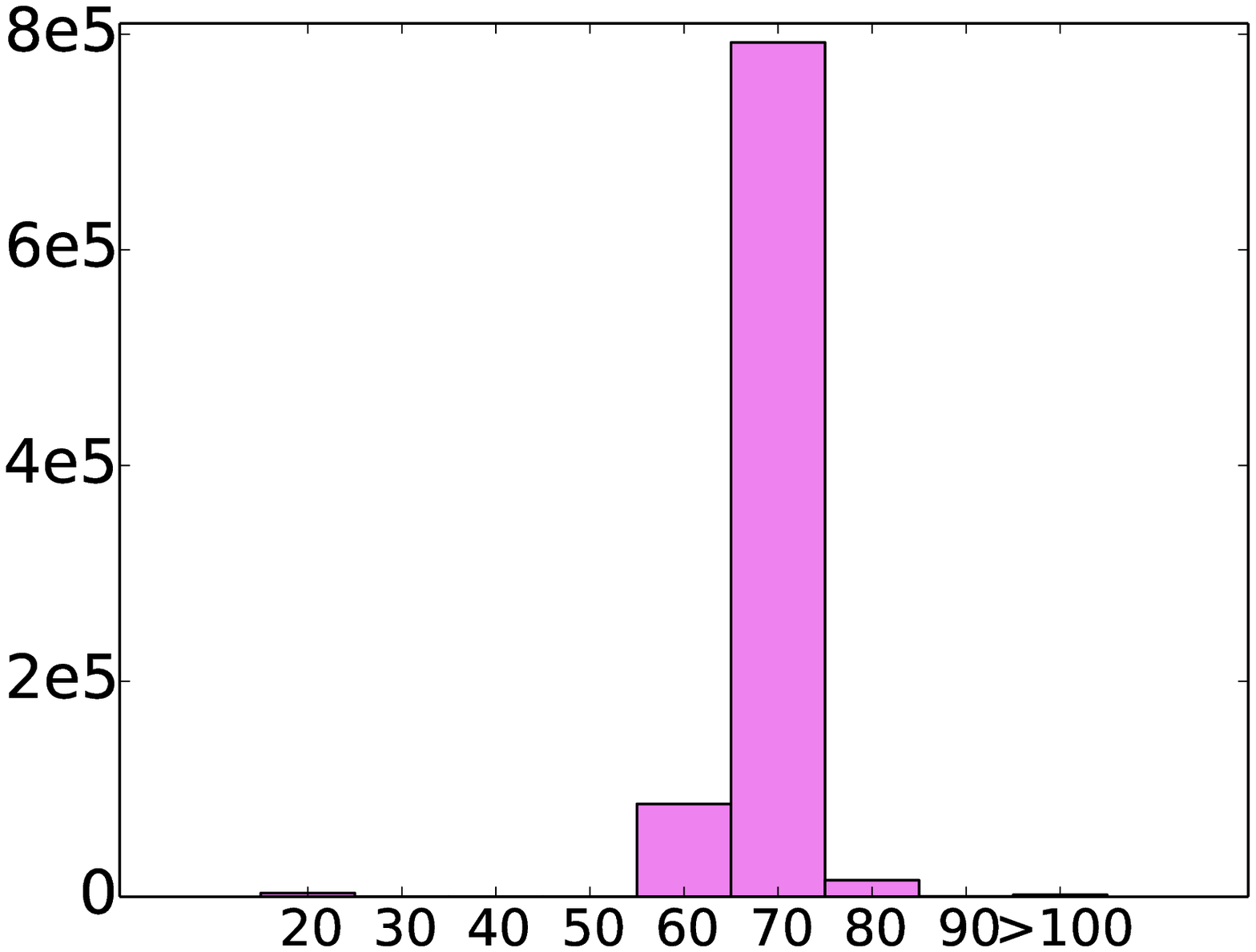}
    \label{fig:prefetch_l3_standard_present}
}
\subfloat[Terminal fault, data in LLC, w/ prefetching.]{
    \includegraphics[clip,width=0.47\columnwidth]{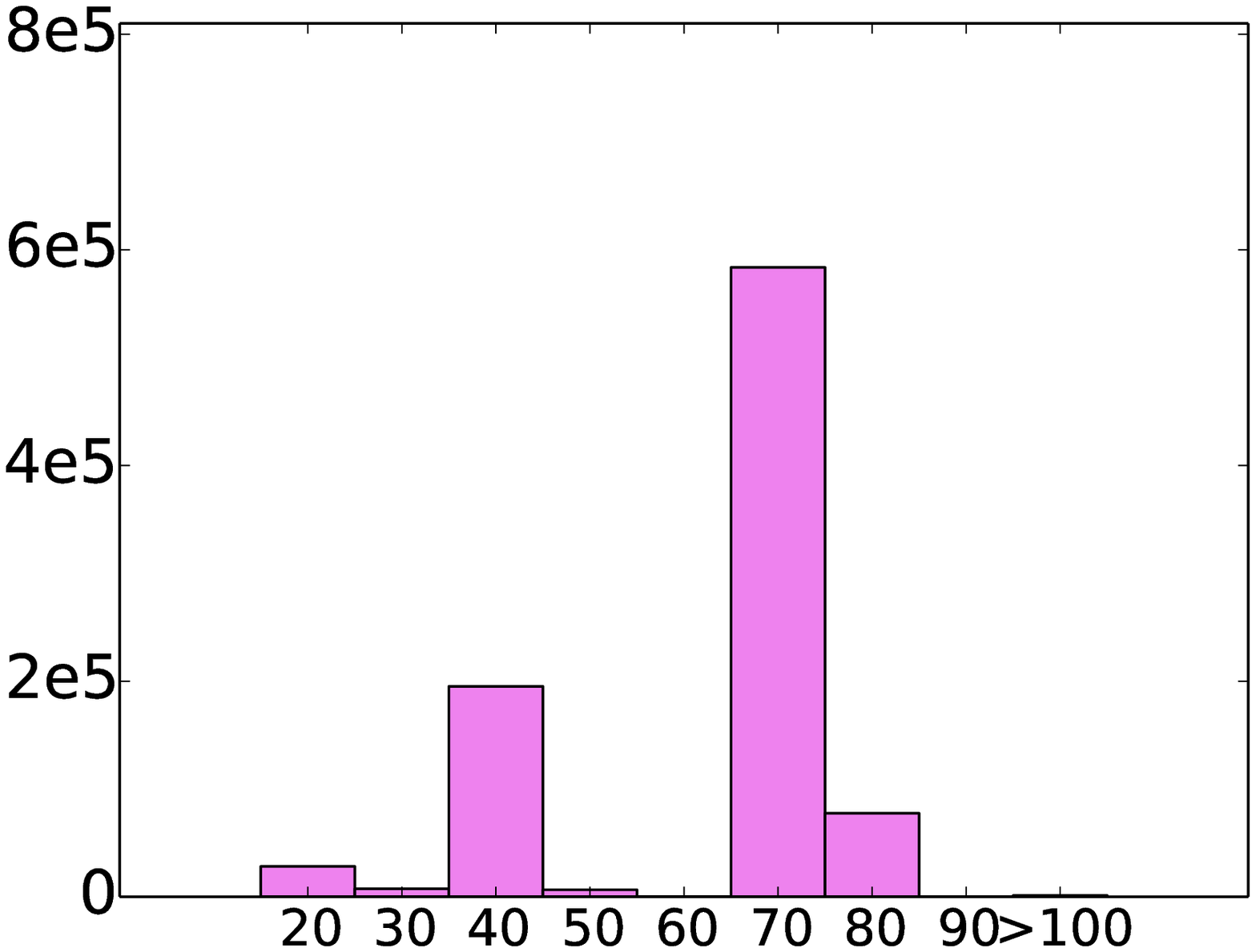}
    \label{fig:prefetch_l3_1000_present}
}
\caption{Reload latency distribution after using Speculation Primitive as a prefetcher.}
\label{fig:prefetch}
\end{figure}

\bheading{Results.}
When the Meltdown-US variant is selected as the Speculation Primitive, the result of the experiment is shown in \figref{fig:prefetch}. In particular, \figref{fig:prefetch_l3_standard} ($N=0$) and \figref{fig:prefetch_l3_1000} ($N=1000$) show the latency of reloading the secret data that is already in the LLC.
Clearly with speculative prefetching, the reload latency drops to the range that is close to L2 cache hit. 
In addition, if a Disclosure Primitive is used to monitor the covert channel while measuring the reload latency, 1/1000 of the time the correct signal can be received while other times a zero signal is. Because in our previous test we have confirmed that only when data is placed in the L1 cache could it be leaked through speculative execution, we conclude that if the data is already in the the L2 cache, the Speculation Primitive has a probability of prefetching it into the L1 cache.
In contrast, in \figref{fig:prefetch_memory_standard}  and \figref{fig:prefetch_memory_1000}, the reload latency distribution does not change with and without prefetching. 

We conducted another experiment to validate our analysis. This time, the Speculation Primitive accesses a memory page with its \texttt{present} flag cleared. We repeated the experiments with the data originally stored in the LLC. The results are shown in \figref{fig:prefetch_l3_standard_present} ($N=0$) and \figref{fig:prefetch_l3_1000_present} ($N=1000$) . Terminal faults have different prefetching effects. After 1000 rounds of Speculation Primitive execution, although some data is preloaded to L2, the peak still keeps at around 70 cycles (LLC). Only zero signals can be received from the covert channel.

\begin{center}
\minibox[frame, rule=1pt,pad=3pt]{

\begin{minipage}[t]{0.95\columnwidth}
\textbf{Conclusion:} In Meltdown-US, if the data is already in the LLC, the Speculation Primitive may prefetch it to the L2 cache and with some probability the L1 cache; if the data is in the memory, the Speculation Primitive cannot prefetch it to the LLC. In Meltdown-P, prefetching with terminal faults is almost impossible to load the data to L1D. 
\end{minipage}
}
\end{center}

\bheading{Misunderstanding regarding Meltdown-US.} Since the publication of the first Meltdown-US paper, there has been a debate whether the attack can be successful if the data is not cached (\ie, stored in the memory). Our work concludes that with only \textit{one round} of Meltdown-US attack, it is only possible to leak the data if it is already in the L1 cache. However, as in most demonstrated Meltdown-US attacks, \textit{multiple rounds} of attack are performed, the secret data can be prefetched to the L1 cache if it is already in the LLC or L2. These findings coincide with our conclusions. Our study provide an explanation for the prior works. In fact, no one has demonstrated Meltdown-US to leak non-cached data---unless the data has been preloaded into the Line Fill Buffer by a thread running in parallel. We have confirmed this fact with the Meltdown-US authors. On the other hand, our study also confirms that L1TF attacks can only succeed when data is already in the L1 cache, the mechanisms of which have not been thoroughly discussed in prior works. 

\subsection{Misprediction Handling and Spectre-type Attacks}
\label{sec:spectre}

\bheading{Branch misprediction handling.}
As an extended study, the two-phase model also applies to Spectre-type misprediction handling. When branch instructions are executed in the back end to determine the branch target address, they leverage the branch prediction units (BPU) to make predictions before the execution completes to improve performance. BPU is a front-end component, making instruction fetching to be immediately redirected to the predicted branch target address. However, if the branch execution unit detects a mismatch between the predicted target address and the true target address, a misprediction is captured by the processor. We expect the misprediction handling is different from exception handling: in \phaseone of misprediction handling, everything required for the handling is performed. The true branch target address is passed to the front end right away once it is determined. All subsequent \uops in ROB are squashed. The IDQ is also flushed. There is no point for the handling to wait until requirement, especially for performance consideration. 

\bheading{Test goals.}
The test aims to verify whether the speculative execution is terminated as soon as the prediction is found to be incorrect, \ie, at \phaseone of the branch instruction, in contrast to at \phasetwo of the exception handling.

\begin{lstlisting}[float=tb,caption={Examing misprediction handling.},label={lst:misprediction}]
 // %RAX: address of uncached memory buffer
 // %RCX: memory address whose data is an address
 // to a covert channel buffer
 // ---------------------------------------------
 // windowing gadget
    [movq (%rax), %rax]
 // ---------------------------------------------
 // speculation primitive
    branch // poisoned to disclosure gadget
    ...
 // ---------------------------------------------
 // disclosure gadget
    movq (%rcx), %rcx
    [add $1, %rcx]
    [sub $1, %rcx]
    ...
    movq (%rcx), %rcx
\end{lstlisting}

\bheading{Experiments.}
It is already known that \phasetwo can be delayed by deferring the retirement of the branch instruction. Thus, the key idea of the experiment is to determine whether the speculative window grows accordingly.

The experiment still follows the basic design of the instruction sequences, but the Speculation Primitive is now a branch instruction. Following the sample code in \lstref{lst:misprediction}, the branch instruction is poisoned to speculatively execute the Disclosure Gadget. The first speculative instruction in the Disclosure Gadget loads a value. The last instruction uses this value as the address of a covert channel buffer. Then \sysname gradually inserts as many ADD/SUB instructions between them as possible for the last instruction to still be executed. The execution of the last instruction is determined using a \flushreload covert channel Disclosure Primitive. The speculative window is measured by counting the maximum number of permitted ADD/SUB instructions.

The experiment requires a differential analysis. In the first run of the test, no Windowing Gadget is implemented. In order to slow down the retirement of the branch instruction, a slow memory load instruction is placed in the Windowing Gadget in the second run for comparison. Given different \phasetwo latency, we observe whether the speculative window are also different.

\bheading{Results.}
Compared to exception handling for Meltdown-type attacks, misprediction handling is known to be exploitable. The only characteristics of interest is the speculation window. During our experiments, it is found that the speculation window remains unchanged with or without the slow Windowing Gadget. Therefore, we conclude that the speculation window of a branch instruction is only determined by \phaseone but not \phasetwo. This is reasonable since branch prediction is designed to optimize performance. Squashing the instructions on the wrong path as soon as possible helps speeding up the execution of the instructions on the correct path. 

\begin{center}
\minibox[frame, rule=1pt,pad=3pt]{

\begin{minipage}[t]{0.95\columnwidth}
\textbf{Conclusion:} Misprediction is handled at \phaseone. Speculative execution stops as soon as misprediction is captured.
\end{minipage}
}
\end{center}
\section{Discussion}

\bheading{Extending \sysname.}
\sysname can be extended to analyze other \vulname variants systematically. For example, the recently disclosed RIDL and Zombieload attacks exploit processor internal buffers (\eg, LFB). Specifically, the Data Accessibility Controller can be extended to control the status of these buffers. We will leave such implementation to future work.
In addition, hardware extensions such as SGX, TSX, and VMX can also be tested by \sysname.  
However, some difficulties may arise: First, it is impossible to identify a comprehensive list of faults for these extensions. Second, some faults may be handled silently, without triggering exceptions. 

\bheading{Limitation.}
There are still limitations to \sysname and we also consider countering them in the future work. 
First, \sysname requires some manual efforts to construct the tests from the exception lists collected from vendor manuals. 
For each exception type in the manual, we still need to determine whether it involves security protection (which is desired) and if so, whether it forms a one-instruction or a two-instruction Speculation Primitive, as demonstrated in \appref{sec:exception_list}. Then based on the conditions to trigger the exception, we need to manually select a Speculation Primitive and develop scripts to automate the tests.
Second, \sysname is unable to perform tests on the variants that trigger micro-architecture events that cannot be unobserved by the Disclosure Gadget (using covert channels). 
Third, \sysname cannot test exceptions not described in the manufacturer's manual.

\section{Related Work}
\label{sec:related}

Closest to our study is by Canella \etal \cite{canella2018systematic} that aims to systematically categorize \vulname attacks. In their work, these attacks are classified into two categories: Meltdown-type \cite{Lipp2018meltdown} and Spectre-type \cite{Kocher2018spectre}. Meltdown-type attacks consider attacks from untrusted but confined programs; Spectre-type attacks assume a benign program tricked to speculatively execute unintended control flows. Both this work and Canella \etal~\cite{canella2018systematic} aim for comprehensive analyzing attack variants. While Canella \etal focuses on attack taxonomization, however, our work emphasizes on the understanding and modeling of fault handling mechanisms. Moreover, a core contribution of our work is the \sysname framework that helps automate testing of such vulnerabilities on commodity processors. Moreover, our work provides insights into the inexploitability of certain vulnerabilities on a particular hardware, bringing a level of assurance to users of these machines.

\bheading{Meltdown-type attacks.}
Meltdown-type attacks can be categorized according to where the secret is stole from. 

\begin{packeditemize}
\item \textit{Memory in separated address spaces.} 
The original Meltdown-US attack \cite{Lipp2018meltdown} leverages out-of-order execution to extract secret data from kernel-space memory. \texttt{L1TF-OS} extracts OS or SMM \cite[Chapter~34]{IntelDevelopmentManual} memory. \texttt{L1TF-VMM} accesses memory of another guest VM or the hypervisor from a non-privileged guest VM. The recently disclosed RIDL \cite{van2019ridl}, Zombieload \cite{schwarz2019zombieload} and Fallout \cite{fallout} are somewhat different from previous attacks since they leverage processor internal buffers as a source of leakage. They leak only in-flight data that are already in these buffers, but meanwhile relax the constraints of address matching. 

\item \textit{Memory in the same address space.}
The Foreshadow attack \cite{vanbulck2018foreshadow} (also called the L1TF attack \cite{intell1tf}) steals secret data from an SGX \cite{intelsgx} enclave from the process that creates the enclave. By clearing the present flag in the PTE of the enclave address, it induces a page fault and thus performs a Meltdown-like attack. \texttt{Meltdown-PK}~\cite{canella2018systematic} reads data speculatively from a memory page protected with Intel protection keys \cite[Chapter~4.6.2]{IntelDevelopmentManual}.
\texttt{Meltdown-BR} \cite{canella2018systematic} bypasses boundary check instructions: When an array in memory is accessed with an index over the bound, the boundary check instructions trigger a range exceeded exception (\#BR). However, it does not prevent the out-of-order execution from accessing the address out of the boundary. Kiriansky \etal \cite{kiriansky2018speculative} exploits speculative writing instead of reading. Memory writes to read-only pages raise an exception but the results could still be speculatively used by following instructions.

\item \textit{Restricted registers.}
LazyFP \cite{stecklina2018lazyfp} exploits lazy FPU context switching to speculatively read register values used before context switches, even though such accesses trigger a device-not-available (\#NM) exception. A Variant 3a disclosed by both ARM \cite{armvariant3a} and Intel \cite{intelvariant3a} is also a Meltdown attack but it targets the privileged system registers such as MSR. However, during our tests, we did not find such exploitable vulnerabilities on tested machines.

\end{packeditemize}


\bheading{Spectre-type attacks.}
Prior studies on Spectre-type attacks can be categorized by the exploited branch prediction units~\cite{canella2018systematic}.

\begin{packeditemize}
\item \textit{Prediction History Table (PHT)}. The original Spectre attack \cite{Kocher2018spectre} poisons the PHT to enable speculative reading of out-of-bound data. NetSpectre \cite{schwarz2018netspectre} extends this local attack to a remote settings. In contrast, Kiriansky \etal \cite{kiriansky2018speculative} demonstrated out-of-bound data writing using similar techniques. O’Keeffe \etal \cite{sgxspectre} poisons PHT to attack SGX enclaves.

\item \textit{Branch Target Buffer (BTB).} Spectre v2 \cite{Kocher2018spectre} targets BTB storing the branch targets of indirect branch instructions. SGXPectre \cite{chen2018sgxpectre} makes use of this variant to steal secret from SGX enclaves.

\item \textit{Return Stack Buffer (RSB)}. Koruyeh \etal \cite{koruyeh2018spectre} and Maisuradze \etal \cite{maisuradze2018ret2spec} demonstrated the poisoning of RSB to trigger speculative side channels.

\item \textit{Store-to-Load Buffer}. Data dependency and data disambiguation related to the Store-to-Load Buffer (although not a prediction unit) were exploited by Horn \cite{variant4} to perform Spectre-type attacks. 

\end{packeditemize}

\section{Conclusion}
\label{sec:conclude}

This paper describes a software framework called \sysname, which enables systematic investigation and quantitative measurement of a variety of \vulname vulnerabilities on commodity processors. We have applied \sysname to test the exploitability of 21 vulnerability variants on 9 processors, confirming prior disclosed vulnerabilities and also uncovering new ones. Moreover, our study explains the root causes of some observations made by prior studies and clarifies common misunderstandings, which paves the paths for future studies.



\bibliographystyle{IEEEtranS}
\bibliography{IEEEabrv,paper}


\appendix

\subsection{Categorized Exception List}
\label{sec:exception_list}

The categorized exception list provided below is summarized from the \textit{Exception and Interrupt Reference}~\cite[Chapter~6.15]{IntelDevelopmentManual}. The * and ** marks refer to one-instruction and two-instruction template respectively as explained in \secref{subsec:instruction_model}. Notice that the original list does not include exceptions or protections from hardware extensions. In our categorized list and in the current work, we do not consider hardware extensions as well. Only some general description about the exceptions triggered by hardware extensions is included in the list below.

\bheading{Page Table Based Data Protection}

\begin{enumerate}
\item Page-Fault Exception (\#PF)
    \begin{itemize}
		\item p bit cleared *[\textit{PTE(Present)}]
		\item user mode accesses
		    \begin{itemize}
			    \item access to supervisor-mode page *[\textit{PTE(US)}]
			    \item write to read-only page **[\textit{PTE(write w/ RW=0)}]
			    \item CR4.PKE = 1, access to user-space page forbidden access by MPK *[\textit{Protection Key (User)}]
	        \end{itemize}
		\item kernel mode accesses
		    \begin{itemize}
			    \item CR4.SMAP = 1, access to user-space page *[\textit{SMAP violation}]
			    \item CR4.PKE = 1, access to user-space page forbidden access by MPK *[\textit{Protection Key (Kernel)}]
			    \item CR0.WP write to read-only page **
	        \end{itemize}
		\item reserved bits not all cleared *[\textit{PTE(Reserved)}]
		\item An enclave access violates one of the specified access-control requirements.
	\end{itemize}
\item Virtualization Exception (\#VE) - EPT violations
\end{enumerate}

\bheading{Segmentation Based Data Protection}

\begin{enumerate}
\item General Protection Exception (\#GP) - Segment-related protection
    \begin{itemize}
        \item Exceeding the segment limit when accessing the CS, DS, ES, FS, or GS segments. *[\textit{DS Over-Limit}]
        \item Loading the DS, ES, FS, or GS register with a segment selector for an execute-only code segment. **[\textit{DS Execute-Only}]
        \item Reading from an execute-only code segment. *[\textit{CS Execute-Only}]
        \item Writing to a code segment or a read-only data segment. **[\textit{DS Read-Only}]
        \item Loading the SS register with a segment selector for a read-only segment. **[\textit{SS Read-Only}]
        \item Accessing memory using the DS, ES, FS, or GS register when it contains a null segment selector. *[\textit{DS Null}]
        \item Loading the SS, DS, ES, FS, or GS register with a segment selector for a system segment. **[\textit{SS $DPL \neq CPL$}]
        \item Transferring execution to a segment that is not executable.
        \item Loading the CS register with a segment selector for a data segment or a null segment selector.
        \item Exceeding the segment limit when referencing a descriptor table (except during a task switch or a stack switch).
        \item Attempting to access an interrupt or exception handler through an interrupt or trap gate from virtual-8086 mode when the handler’s code segment DPL is greater than 0.
    \end{itemize}
\item Data Segment Not Present (\#NP) *[\textit{DS Not-Present}]
\item Stack Fault Exception (\#SS)
    \begin{itemize}
        \item Limit violation when accessing ss register (eg. pop) *[\textit{SS Over-Limit}]
        \item Loading non-present stack into SS register. **[\textit{SS Not-Present}]
        \item Loading the SS register with the segment selector of an executable segment or a null segment selector. **[\textit{SS Null}]
    \end{itemize}
\end{enumerate}

\bheading{Program Instruction Based Data Protection}

\begin{itemize}
\item BOUND Range Exceeded Exception (\#BR) **[\textit{BOUND}]
\item Intel MPX
\end{itemize}

\bheading{Other Protection}

\begin{itemize}
\item Device Not Available Exception (\#NM) - Lazy context save after context switch (CR0.TS) *[\textit{Load xmm0 (CR0.TS)}]
\item SMM memory access protection
\item General Protection Exception (\#GP)
    \begin{itemize}
    \item Attempting to execute a privileged instruction when the CPL is not equal to 0. (MOV (load) control/debug registers, RDMSR) *[\textit{Load CR4}] \& *[\textit{Load MSR (0x1a2)}]
    \item Attempting to execute SGDT, SIDT, SLDT, SMSW, or STR when CR4.UMIP = 1 and the CPL is not equal to 0.
    \item Attempting to execute a privileged (serializing) instruction when the CPL is not equal to 0 (LGDT, LLDT, LTR, LIDT, MOV [store] (control registers / debug registers), LMSW, CLTS, WRMSR).
    \item Executing the INT n instruction when the CPL is greater than the DPL of the referenced interrupt, trap, or task gate.
    \end{itemize}
\end{itemize}

\bheading{Arithmetic Protection}

\begin{itemize}
\item Overflow Exception (\#OF) - INTO instruction
\item x87 FPU Floating-Point Error (\#MF)
\item SIMD Floating-Point Exception (\#XM)
\item Divide Error Exception (\#DE)
\end{itemize}

\bheading{Non Protection}

\begin{itemize}
\item Debug Exception (\#DB)
\item Breakpoint Exception (\#BP) - INT3 instruction
\item Invalid Opcode Exception (\#UD)
\item Double Fault Exception (\#DF)
\item Invalid TSS Exception (\#TS)
\item General Protection Exception (\#GP)
    \begin{itemize}
        \item Accessing a gate that contains a null segment selector.
        \item The segment selector in a call, interrupt, or trap gate does not point to a code segment.
        \item The segment selector operand in the LLDT instruction is a local type (TI flag is set) or does not point to a segment descriptor of the LDT type.
        \item The segment selector operand in the LTR instruction is local or points to a TSS that is not available.
        \item The target code-segment selector for a call, jump, or return is null.
        \item Using a segment selector on a non-IRET task switch that points to a TSS descriptor in the current LDT. TSS descriptors can only reside in the GDT. This condition causes a \#TS exception during an IRET task switch.
        \item Instruction length limit exceeded.
        \item Loading CR0 with PG=1 (paging enabled) and PE=0 (protection disabled). / Loading CR0 with NW=1 and CD=0.
        \item Attempting to write a 1 into a reserved bit of CR4/MSR/MXCSR/(64-bit)CR3, CR4 or CR8.
        \item If the PAE and/or PSE flag in control register CR4 is set and the processor detects any reserved bits in a page-directory-pointer-table entry set to 1.
        \item Referencing an entry in the IDT (following an interrupt or exception) that is not an interrupt, trap, or task gate.
        \item (64-bit) Non-canonical address / null address memory access.
        \item Executing an SSE/SSE2/SSE3 instruction that attempts to access a 128-bit memory location that is not aligned on a 16-byte boundary when the instruction requires 16-byte alignment. This condition also applies to the stack segment.
        \item An attempt is made to clear CR0.PG while IA-32e mode is enabled.
    \end{itemize}
\item Alignment Check Exception (\#AC)
\item Machine-Check Exception (\#MC)
\item Stack Fault Exception (\#SS)
    \begin{itemize}
        \item There is not enough stack space for allocating local variables when executing ENTER instruction.
        \item (64-bit) Non-canonical address using SS register.
    \end{itemize}
\end{itemize}

\subsection{Choosing The Best Suppressing Primitive} 

\begin{figure}[t]

    \subfloat[Exception-based suppression.]{
        \includegraphics[clip,width=0.47\linewidth]{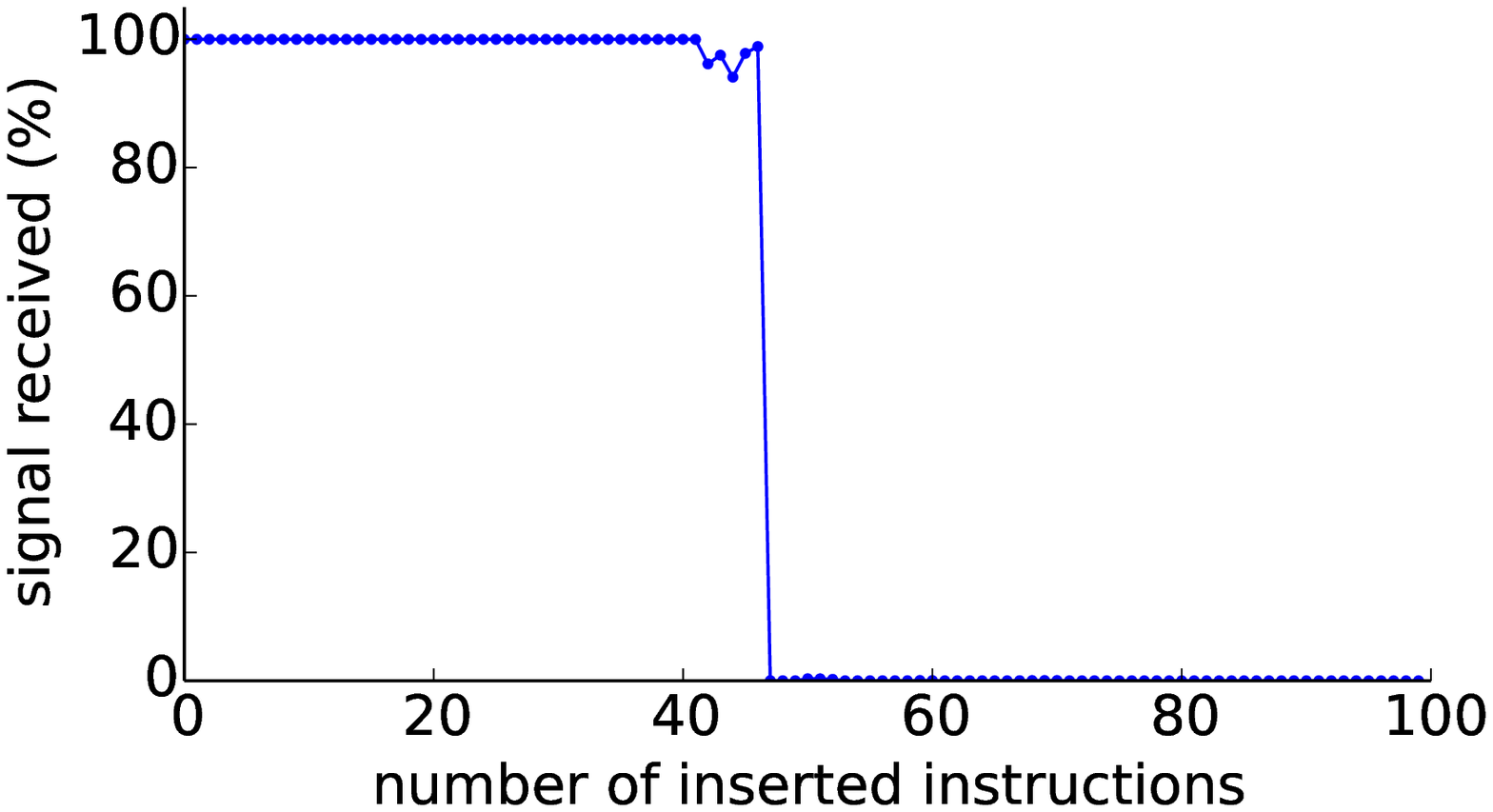}
        \label{fig:p1_exception}
    }
    \subfloat[Retpoline-based suppression]{
        \includegraphics[clip,width=0.47\linewidth]{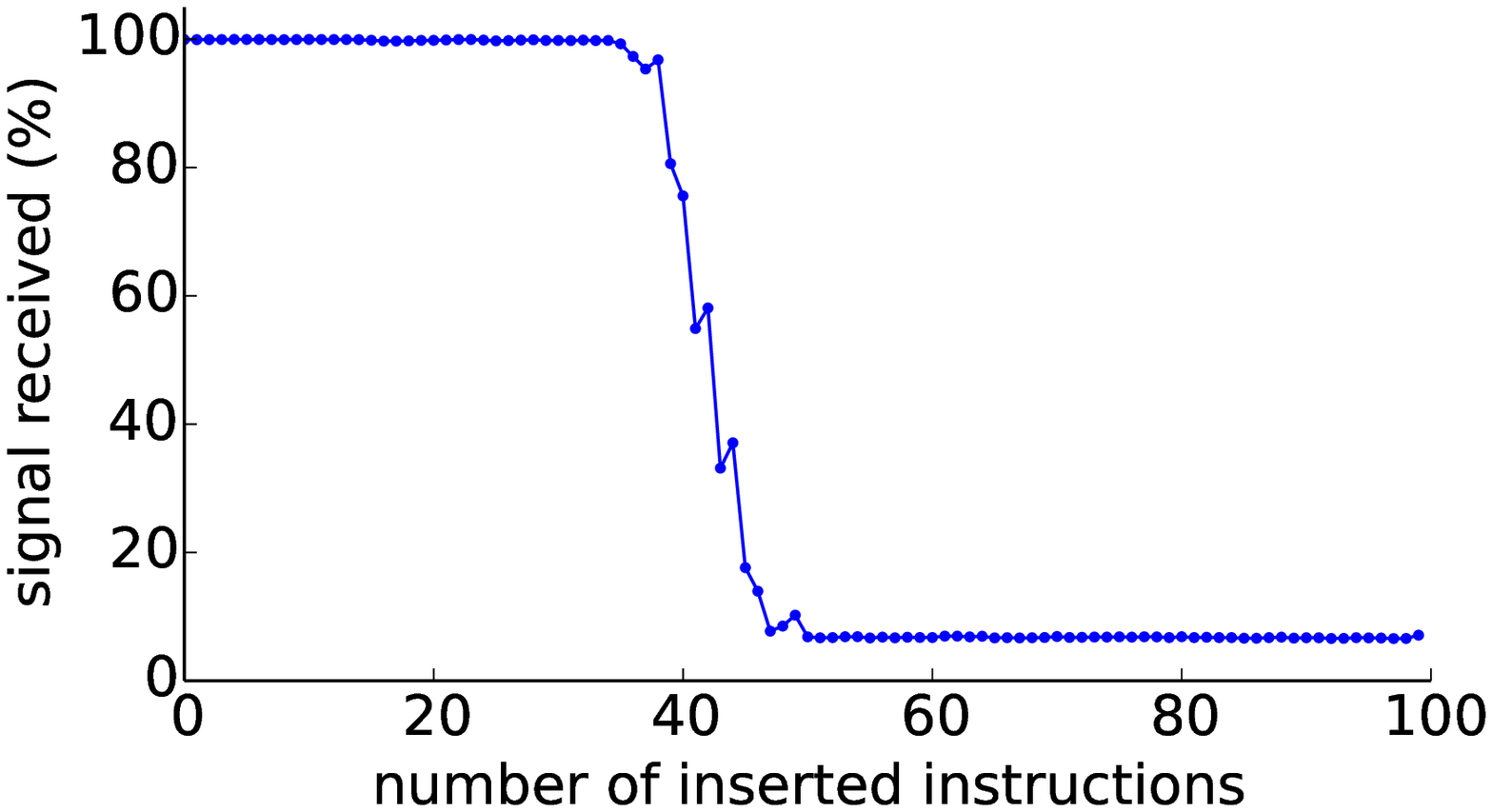}
        \label{fig:p1_retpoline}
    }
    \caption{Comparison between different suppression methods.}

\end{figure}

We explore two implementations of the Suppressing Primitive: exception-based and retpoline-based. An exception-based Suppressing Primitive is illustrated in \lstref{lst:framework2}. The other uses a retpoline to suppress exceptions~\cite{stecklina2018lazyfp}. We evaluated these methods using the following method: 
Using each method, we gradually insert ADD/SUB instructions to find the maximum speculation window. The covert-channel tests were repeated 100,000 times for each number of inserted instructions. The rate of receiving the covert-channel signals for each number is illustrated in  \figref{fig:p1_exception} and \figref{fig:p1_retpoline}, respectively. It can be seen from the figures that the retpoline-based approach is less desirable as the rate of receiving the signal drops gradually when the inserted instructions increases, making it hard to determine the speculation window. Therefore, in our test, the exception-based approach is used. 

\end{document}